\documentclass[sigconf,screen,natbib=false,nonacm]{acmart}
\acmConference[ASE 2022]{ASE '22: Proceedings of the 37th IEEE/ACM International Conference on Automated Software Engineering}{October 10-14, 2022}{Ann Arbor, MI, USA}

\usepackage[utf8]{inputenc}

\usepackage[style=ACM-Reference-Format,backend=biber,defernumbers=true,sortcites=true]{biblatex}

\makeatletter
\let\blx@rerun@biber\relax
\makeatother

\usepackage{subcaption}
\usepackage{multirow}
\usepackage{xspace}
\usepackage{listings}
\usepackage{soul}
\usepackage{colortbl}
\usepackage{hyperref}

\newcommand{\canvas}{\texttt{<canvas>}\xspace}
\newcommand{\imgtag}{\texttt{<img>}\xspace}

\begin{document}
\title{Automatically Detecting Visual Bugs in HTML5 \texttt{<canvas>} Games}
\author{Finlay Macklon}
\email{macklon@ualberta.ca}
\affiliation{
  \institution{University of Alberta}
  \city{Edmonton}
  \state{AB}
  \country{Canada}
}
\author{Mohammad Reza Taesiri}
\email{taesiri@ualberta.ca}
\affiliation{
  \institution{University of Alberta}
  \city{Edmonton}
  \state{AB}
  \country{Canada}
}
\author{Markos Viggiato}
\email{viggiato@ualberta.ca}
\affiliation{
  \institution{University of Alberta}
  \city{Edmonton}
  \state{AB}
  \country{Canada}
}
\author{Stefan Antoszko}
\email{santoszk@ualberta.ca}
\affiliation{
  \institution{University of Alberta}
  \city{Edmonton}
  \state{AB}
  \country{Canada}
}
\author{Natalia Romanova}
\email{natalia.romanova@prodigygame.com}
\affiliation{
  \institution{Prodigy Education}
  \city{Toronto}
  \state{ON}
  \country{Canada}
}
\author{Dale Paas}
\email{dale.paas@prodigygame.com}
\affiliation{
  \institution{Prodigy Education}
  \city{Toronto}
  \state{ON}
  \country{Canada}
}
\author{Cor-Paul Bezemer}
\email{bezemer@ualberta.ca}
\affiliation{
  \institution{University of Alberta}
  \city{Edmonton}
  \state{AB}
  \country{Canada}
}
\date{May 2022}

\begin{abstract}
	The HTML5 \texttt{<canvas>} is used to display high quality graphics in web applications such as web games (i.e., \texttt{<canvas>} games). However, automatically testing \texttt{<canvas>} games is  not possible with existing web testing techniques and tools, and manual testing is laborious. Many widely used web testing tools rely on the Document Object Model (DOM) to drive web test automation, but the contents of the \texttt{<canvas>} are not represented in the DOM. The main alternative approach, snapshot testing, involves comparing oracle snapshot images with test-time snapshot images using an image similarity metric to catch visual bugs, i.e., bugs in the graphics of the web application. However, creating and maintaining oracle snapshot images for \texttt{<canvas>} games is onerous, defeating the purpose of test automation. In this paper, we present a novel approach to automatically detect visual bugs in \texttt{<canvas>} games. By leveraging an internal representation of objects on the \texttt{<canvas>}, we decompose snapshot images into a set of object images, each of which is compared with a respective oracle asset (e.g., a sprite) using four similarity metrics: percentage overlap, mean squared error, structural similarity, and embedding similarity. We evaluate our approach by injecting 24 visual bugs into a custom \texttt{<canvas>} game, and find that our approach achieves an accuracy of 100\%, compared to an accuracy of 44.6\% with traditional snapshot testing.

\end{abstract}

    
\maketitle

\section{Introduction}\label{sec:introduction}
\begin{figure}[t]
\centering
\setlength{\fboxsep}{0pt}%
\begin{subfigure}[t]{\linewidth}
\centering
\fbox{\includegraphics[width=\textwidth]{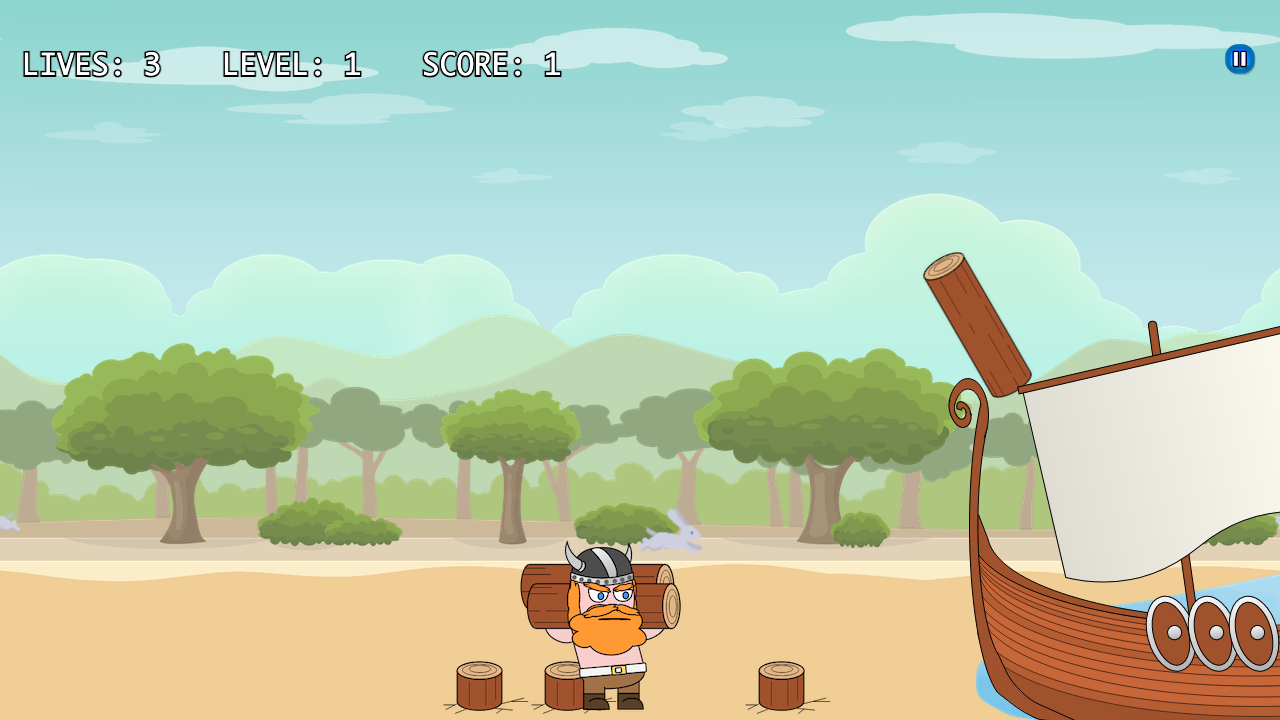}}
\caption{Screenshot of our test \canvas game}
\label{fig:sampleframe}
\end{subfigure}
\hfill
\begin{subfigure}[t]{0.975\linewidth}
\begin{lstlisting}[language=html,basicstyle=\ttfamily\footnotesize,frame=single]
<!DOCTYPE html>
<html>
<head>
    <style> body { margin: 0; display: flex; } </style>
    <script src="main.js"></script>
</head>
<body>
    <canvas width="1280px" height="720px"></canvas>
</body>
</html>
\end{lstlisting} 
\caption{HTML code for our test \canvas game}
\end{subfigure}
\caption{The graphics of a \canvas game are represented as a bitmap, and not in the DOM, while the game's source code resides in the script \texttt{main.js}.}
\label{fig:intro}
\end{figure}

The HTML \canvas is used to display high-quality graphics in web applications, and is particularly useful for web games (i.e., \canvas games)~\cite{parisi2012webgl, parisi2014programming, konstantinidis2016moving, yogya2014comparison, fulton2013html5}.
HTML5 \canvas games are receiving growing attention from industry~\cite{google2020gamesnacks, goodboy2020playco}, but it is challenging to automatically test \canvas games, as widely used web testing techniques and tools do not work for the \canvas~\cite{macklon2022taxonomy}.
Many commonly used web testing tools leverage the Document Object Model (DOM) to drive test automation, but as demonstrated in Figure~\ref{fig:intro}, \canvas graphics are not represented in the DOM.

To overcome this challenge, snapshot testing has become the industry standard approach to visual testing for \canvas applications, as it does not rely on the DOM, but instead relies on screenshots of the web application.
Snapshot testing targets visual bugs, i.e., bugs that are related to the graphics of the application, by automatically comparing oracle screenshots with screenshots that are recorded during the execution of a test case.
However, as we discuss in Section~\ref{subsec:snapshottesting}, snapshot testing cannot deal with the dynamic nature of \canvas games, as this dynamism causes variation between the screenshots that is hard to account for automatically. 

Because of the technical challenges of \canvas testing, and the inherent difficulties of testing games~\cite{politowski2020dataset, murphy2014cowboys, stacey2009temporal, petrillo2009went, lewis2010went, kamienski2021empirical}, \canvas games are mostly tested manually.
Manual testing requires large amounts of manual time and effort, limiting the amount of bugs that quality assurance (QA) analysts can realistically discover and report.

\begin{figure*}[t]
	\centering
	\setlength{\fboxsep}{0pt}%
	\begin{subfigure}[t]{0.3\linewidth}
		\centering
		\fbox{\includegraphics[width=\textwidth]{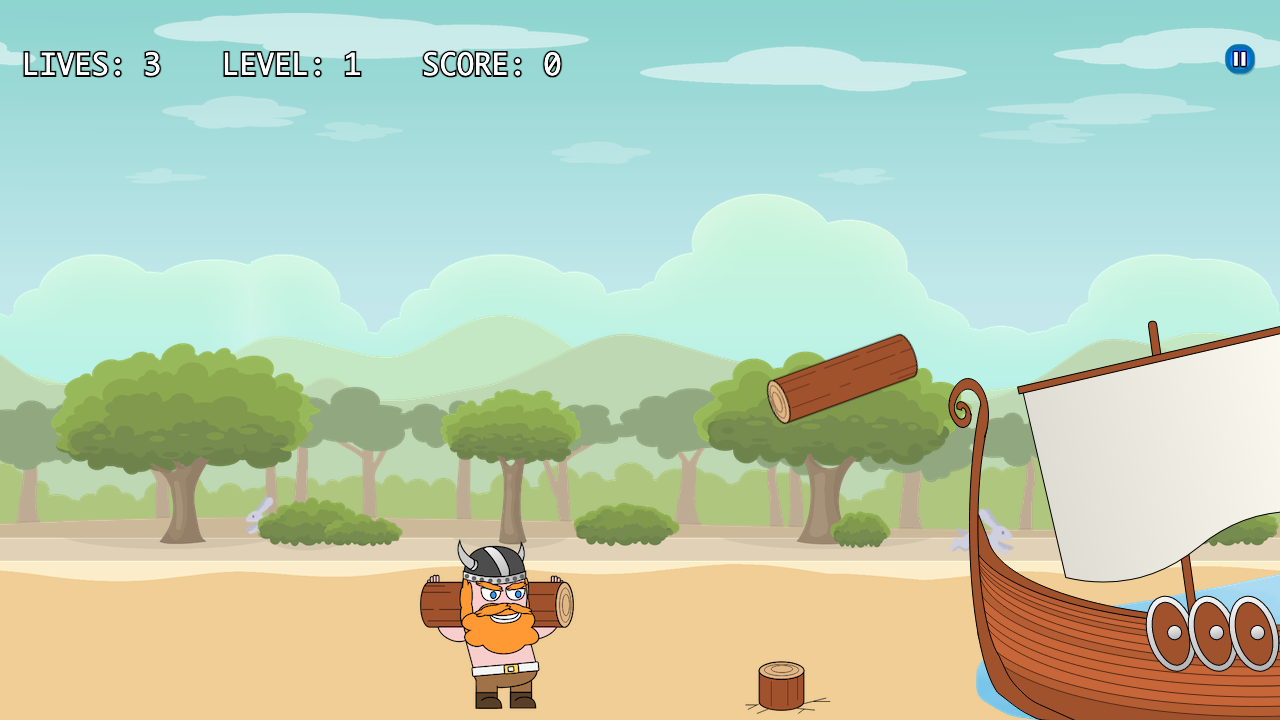}}
		\caption{Oracle screenshot}
		\label{fig:intro_a}
	\end{subfigure}
	\hfill
	\begin{subfigure}[t]{0.3\linewidth}
		\centering
		\fbox{\includegraphics[width=\textwidth]{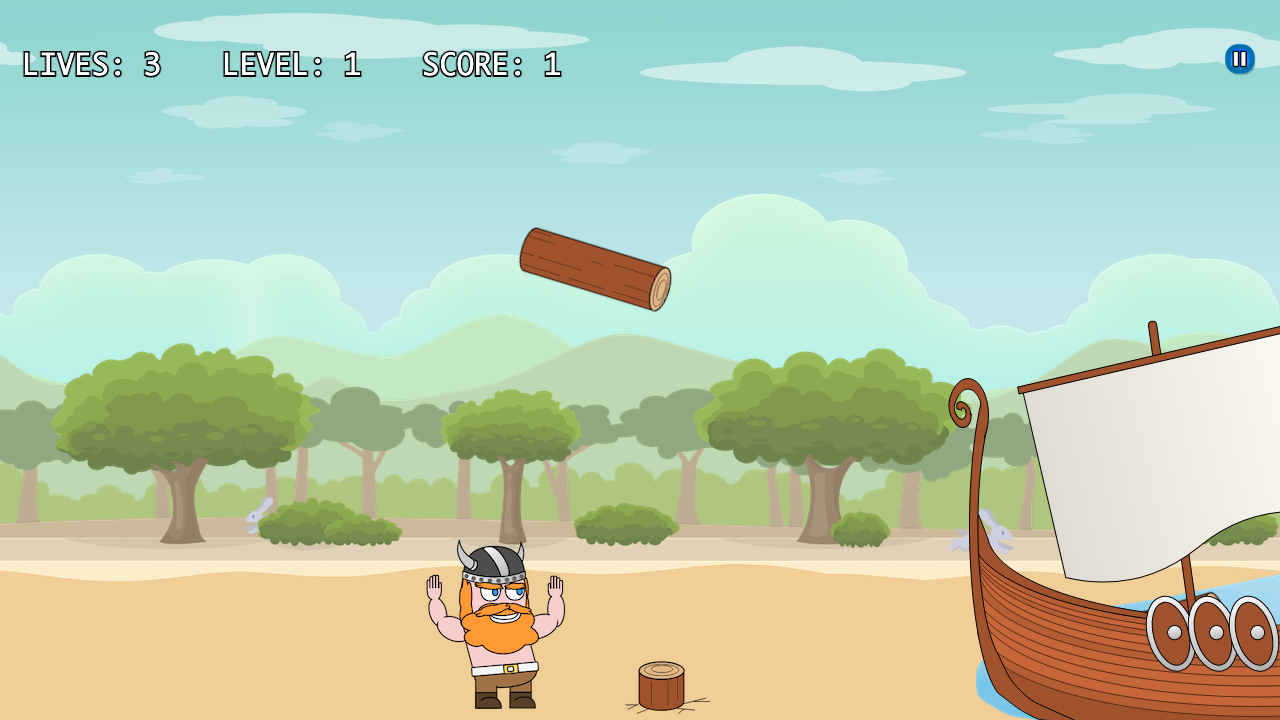}}
		\caption{Test screenshot}
		\label{fig:intro_b}
	\end{subfigure}
	\hfill
	\begin{subfigure}[t]{0.3\linewidth}
		\centering
		\fbox{\includegraphics[width=\textwidth]{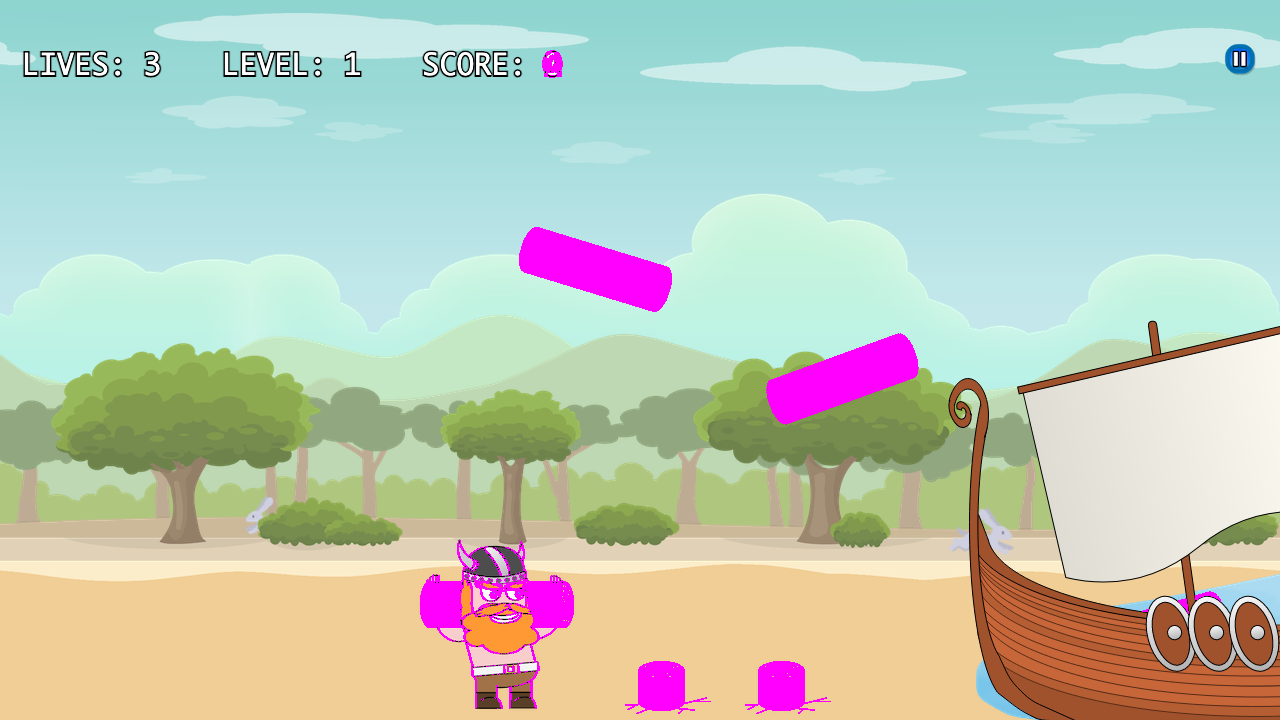}}
		\caption{Image difference (in pink)}
		\label{fig:intro_compare_bug}
	\end{subfigure}
	\caption{Two screenshots from our test game. In the test screenshot, the viking character is missing a log on his shoulders (injected visual bug \hyperref[tab:bugdetection]{\texttt{S4}} in Table~\ref{tab:bugdetection}, \textit{viking animation not updating}). However, as observed in the third screenshot, the in-game randomness causes a larger difference between the screenshots than the bug itself.}
	\label{fig:intro_sample}
\end{figure*}

Therefore, we propose an automated approach for the visual testing of \canvas games.
Rather than (manually) curating oracle screenshots, we use game assets (see Section~\ref{subsec:canvasgames}) to automatically generate visual test oracles during the test, and automatically compare these oracle assets with individual objects on a screenshot of the \canvas.
Our approach leverages the game's internal representation of objects on the \canvas, i.e., the \canvas objects representation (COR), to decompose screenshots of the \canvas into individual object images.

We evaluated our approach by injecting 24 unique visual bugs from 4 different bug types (state, appearance, layout, rendering) as defined by Macklon et al.~\cite{macklon2022taxonomy} into a custom \canvas test game.
Our approach performed automated visual comparisons of the oracle assets and the rendered objects using four similarity metrics: percentage overlap, mean squared error, structural similarity, and embedding similarity.
We compared our approach with a baseline approach that is the industry standard, i.e., snapshot testing.
We found that when using mean squared error, structural similarity, or embedding similarity as the similarity metric, our approach achieves an accuracy of 100\% for the 24 injected bugs in our test game, compared to an accuracy of 44.6\% with the baseline approach.

The main contributions of our paper are as follows:
\begin{itemize}
    \item We designed 24 synthetic visual bugs to evaluate automated testing approaches for \canvas games, and we confirmed with an industrial partner that these bugs were representative of the bugs found in real \canvas games.
    \item We created a testbed for evaluating visual testing techniques for \canvas games, i.e., a test \canvas game, which includes a non-buggy version and a buggy version of the game containing the 24 synthetic bugs.
    \item We extensively evaluated combinations of four widely-used similarity metrics for automatically detecting visual bugs in \canvas games.
    \item For reproducibility, we open-sourced our testbed and visual bugs dataset at the following link: \url{https://github.com/asgaardlab/canvas-visual-bugs-testbed}.
    \item A live version of our test \canvas game is available at the following link:
    \url{https://asgaardlab.github.io/canvas-visual-bugs-testbed/game}.
\end{itemize}

The remainder of our paper is structured as follows.
Section~\ref{sec:background} discusses background information.
Section~\ref{sec:relatedwork} discusses related work.
Section~\ref{sec:ourapproach} presents our approach.
Section~\ref{sec:experiments} details our experiment setup.
Section~\ref{sec:results} presents our results.
Section~\ref{sec:threats} contains threats to validity.
Section~\ref{sec:conclusion} is the conclusion to our paper.

\section{Background}\label{sec:background}
In this section, we give background information about HTML5 \canvas games and snapshot testing. 

\subsection{HTML5 \canvas games}\label{subsec:canvasgames}
By combining the high-quality graphics of the \canvas with browser events, such as mouse clicks, game developers can create complete games that run in a web browser.

\subsubsection*{Open-source frameworks}
It is difficult to integrate the \canvas with other parts of a web application~\cite{macklon2022taxonomy}, and so \canvas \emph{frameworks} are used to ease the development of \canvas games.
There exist several free and open-source (FOSS) \canvas frameworks that are widely-used to develop \canvas games. 
For example, \texttt{PixiJS} and \texttt{Phaser} receive much attention from game developers, as indicated by the high number of forum posts related to each framework on the HTML5 Game Devs~\cite{html5gamedevs:0} and Stack Overflow~\cite{almansoury2020investigating} forums.
Such \canvas frameworks typically provide a custom internal representation of objects on the \canvas, i.e., a \canvas objects representation (COR), which can be manipulated by developers to easily create animations on the \canvas.
For example, in \texttt{PixiJS}, the COR is termed \emph{scene graph}, and has a tree structure.

\subsubsection*{Assets}
A common way to integrate graphics into a video game is using source images (\emph{assets}) that are used to display objects in the game.
For \canvas games, assets are loaded by the web application client from some file server through web requests, like any other image in a web application.
However, assets are not rendered as image (\imgtag) elements on a web page, but instead are used as source bitmaps that are displayed on the \canvas bitmap.

\subsection{Snapshot testing} \label{subsec:snapshottesting}
Snapshot testing, e.g., using \texttt{Percy}, is the industry standard for visually testing web applications~\cite{ricca2021ai}.
Visual testing is used to target visual bugs; visual bugs are mismatches between actual and expected visual properties in the graphics of a software application~\cite{issa2012visual}.
Traditional snapshot testing typically involves comparing screenshots of the web application from the same test across different runs, after some change(s) to the source code (e.g., a pull request).
To perform traditional snapshot testing, first a set of oracle screenshots that have been collected during a test run must be manually verified, and then new test screenshots can be automatically collected and compared at a later time using an image comparison algorithm.
If a screenshot does not pass the image comparison check, that screenshot (or test case) is flagged for manual review.

Figure~\ref{fig:intro_sample} shows how most of the visual differences between the oracle and test screenshots occur due to random elements of the \canvas game, which are desired functionality, rather than the injected visual bug.
It is difficult to distinguish between visual bugs and intended functionality for \canvas games when using snapshot testing.
This problem can lead to many false positives, increasing the manual workload (due to oracle re-verification) and reducing the benefit of using snapshot testing as an automated testing approach.
Therefore, the industry-standard approach for snapshot testing is far from ideal for testing many \canvas applications, particularly \canvas games.

\section{Related Work}\label{sec:relatedwork}
In this section, we discuss related work on \canvas testing, visual web and GUI testing, and visual game testing. 

\subsection{\canvas testing}
Macklon et al.~\cite{macklon2022taxonomy} analyzed open source projects on GitHub that utilize the \canvas, and proposed a taxonomy of \canvas bugs.
They showed that the most frequently reported bugs in the open source projects are visual bugs, i.e., bugs that are related to the graphics of an application.
Their findings emphasize that research on \canvas testing is at an early stage and has many opportunities, and that visual bugs are a primary concern for \canvas testing.

Only one prior study has investigated testing methods for the \canvas. 
Bajammal and Mesbah propose an approach to enable DOM-based testing of the \canvas by leveraging traditional computer vision techniques to detect objects on the \canvas, and subsequently augment the DOM with a representation of those objects~\cite{bajammal2018web}.
They report high accuracy in detecting objects on the \canvas that should not be present (similar to visual bug \hyperref[tab:bugdetection]{\texttt{S6}} in Table~\ref{tab:bugdetection}), however any other type of overlapping visual bug on the \canvas would pose challenges for their visual inference algorithm.
In contrast, we evaluate our approach on 24 unique bugs from 4 visual bugs types, and find that our approach shows strong performance for catching a wide variety of visual bugs that are representative of bugs found in real-world \canvas games.

\subsection{Visual web and GUI testing}
As previously outlined, existing automated web testing techniques and tools do not work for the \canvas, but prior research has also indicated that \canvas bugs overlap with visual bugs found in graphical user interfaces (GUIs) and generic web applications~\cite{macklon2022taxonomy}. 
We refer to the survey of computer vision applications in software engineering by Bajammal et al.~\cite{bajammal2020survey} and the grey literature review of AI-based test automation techniques by Ricca et al.~\cite{ricca2021ai} for an overview of visual testing for GUIs and web apps.
Here, we only discuss related work that was not covered in the survey by Bajammal et al.~\cite{bajammal2020survey}.

Several prior studies have proposed the use of visual analysis to assist in automated testing methods for web applications.
Yandrapally and Mesbah~\cite{yandrapally2021fragment} proposed a method to automatically detect near-duplicate states in web applications by comparing fragments of a web page instead of entire screenshots.
They decomposed the DOM along with screenshots and performed automatic structural and visual comparisons between automatically inferred web page states.
Bajammal and Mesbah~\cite{bajammal2021semantic} automatically inferred the semantic role of regions in a web page and automated the testing of web accessibility requirements.
In another work by Bajammal and Mesbah~\cite{bajammal2021page}, they combined visual analysis with DOM attributes to improve automated web page segmentation, which can assist with bug localization.
These works focus on segmenting and testing the structure of web pages, i.e., what is represented in the DOM, but as previously explained, the contents of the \canvas are not represented in the DOM, meaning these approaches cannot be used to automatically catch visual bugs in \canvas games.

Several prior studies proposed the use of computer vision to leverage the visual aspect of a software application in an automated testing process.
Mazinanian et al.~\cite{mazinanian2021style} automatically predicted actionable elements on a web page through a supervised deep learning approach.
White et al.~\cite{white2019improving} proposed a supervised deep learning approach and automatically identified GUI components to improve the coverage of random testing.
Xue et al.~\cite{xue2022learning}  proposed a supervised deep learning approach to assist in performing record-and-replay GUI testing in a mobile or web application.
Mozgovoy and Pyshkin~\cite{mozgovoy2017unity} used template matching to recognize objects and GUI elements in a screenshot of a mobile game, which allow test assertions to be made against the visual content of the game.
Ye et al.~\cite{ye2021empirical} proposed a similar GUI widget detection approach for mobile games, in which they instrumented the source code of a mobile game to automatically extract samples of GUI widgets, and subsequently trained a supervised deep learning model for GUI widget detection.
Visual bugs would interfere with the GUI element identification methods in the aforementioned works, while our approach instead targets visual bugs in \canvas games without training any new models.

Zhao et al. proposed the use of unsupervised deep learning methods to detect anomalous GUI animations, which requires only several ground truth samples of a correct GUI animation to detect the anomalous animations~\cite{zhao2020seenomaly}.
Given the dynamic nature of \canvas games, it would be extremely challenging to collect ground truth samples of all correct animations in a \canvas game, which does not solve the problems posed by snapshot testing.
We avoid this problem in our approach by automatically generating visual test oracles during the test.

\subsection{Visual game testing}
Given that the \canvas is often used to build web games, we provide an overview of visual testing in video games.

Automated methods for graphics glitch detection in video games have been proposed in prior work. 
Nantes et al. propose a semi-automated approach to detect shadow glitches in a video game using traditional computer vision techniques~\cite{nantes2008framework}.
However, in our work we propose a fully automated approach to detect a wider range of visual bugs that are relevant to \canvas games.

Other studies have utilized relatively recent advancements in deep learning to detect graphics glitches in video games~\cite{davarmanesh2020automating, ling2020using, chen2021glib} or to leverage the visual aspect of video games for sprite and event extraction~\cite{kim2020synthesizing, smirnov2021marionette, luo2018player, luo2019making, taesiri2022clip}.
However, these methods all require either significant manual effort to prepare the data or in-house machine learning expertise to train and fine-tune the models, or they target only a limited set of visual bugs (when compared to the four types evaluated in our paper).
As our approach utilizes a pre-trained model, it requires only very basic applied machine learning knowledge, and it does not require much data preparation.
\section{Our Approach}\label{sec:ourapproach}
\begin{figure}[!t]
    \centering
    \includegraphics[width=\linewidth]{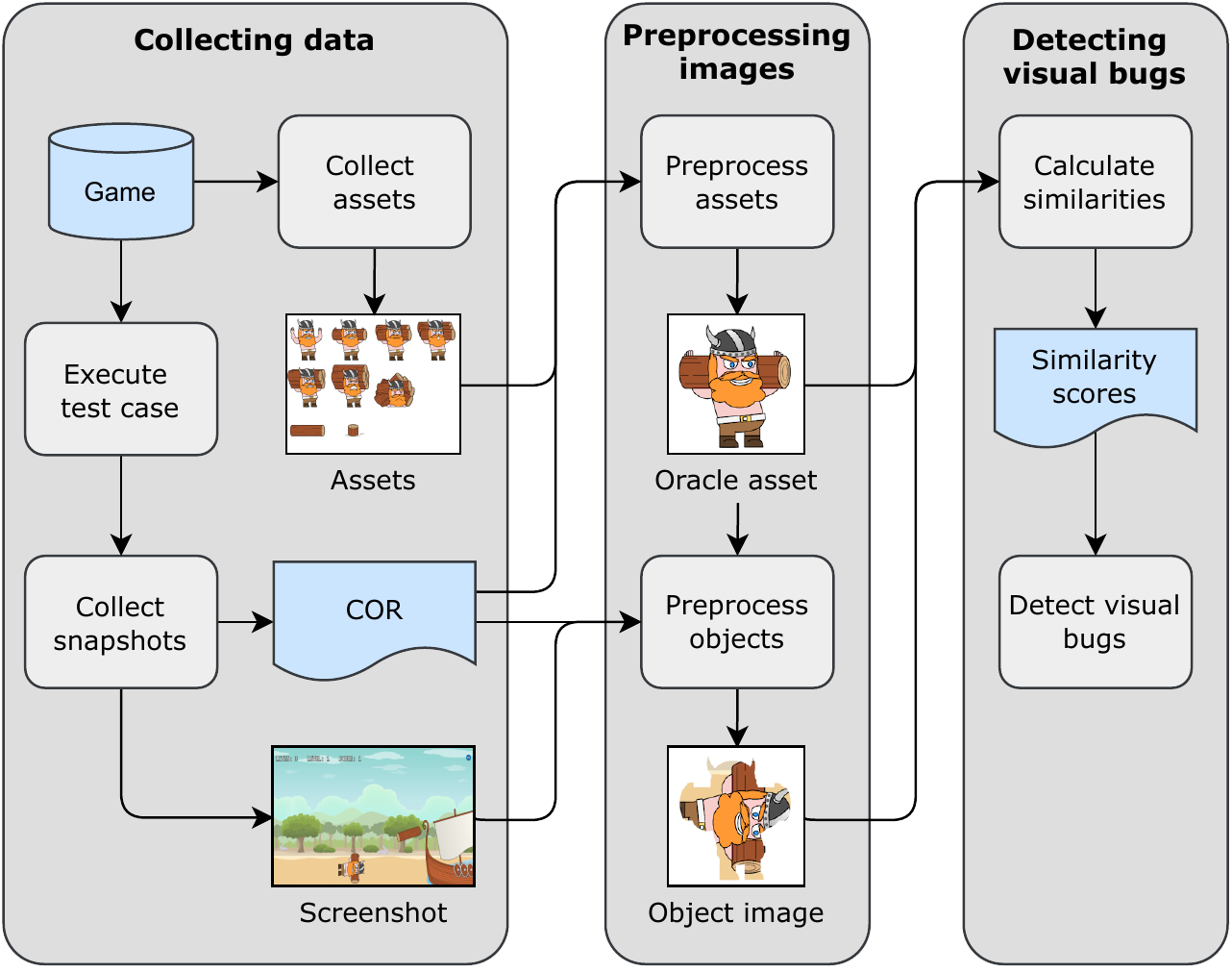}
    \caption{Overview of our approach (shown with visual bug \hyperref[tab:bugdetection]{\texttt{L4}} in Table~\ref{tab:bugdetection}, \emph{Viking has wrong rotation}).}
    \label{fig:overview}
\end{figure}

In this section, we present our approach for automatically detecting visual bugs in \canvas games.
Figure~\ref{fig:overview} shows an overview of the steps of our approach.

\subsection{Collecting data}\label{subsec:collectingdata}
We begin by automatically instrumenting the rendering loop of the \canvas game with our custom code to collect snapshots and assets.
Each snapshot contains a screenshot of the \canvas and a respective \canvas objects representation (COR) from the same point in time.

\begin{figure*}[!t]
    \centering
    \includegraphics[width=\linewidth]{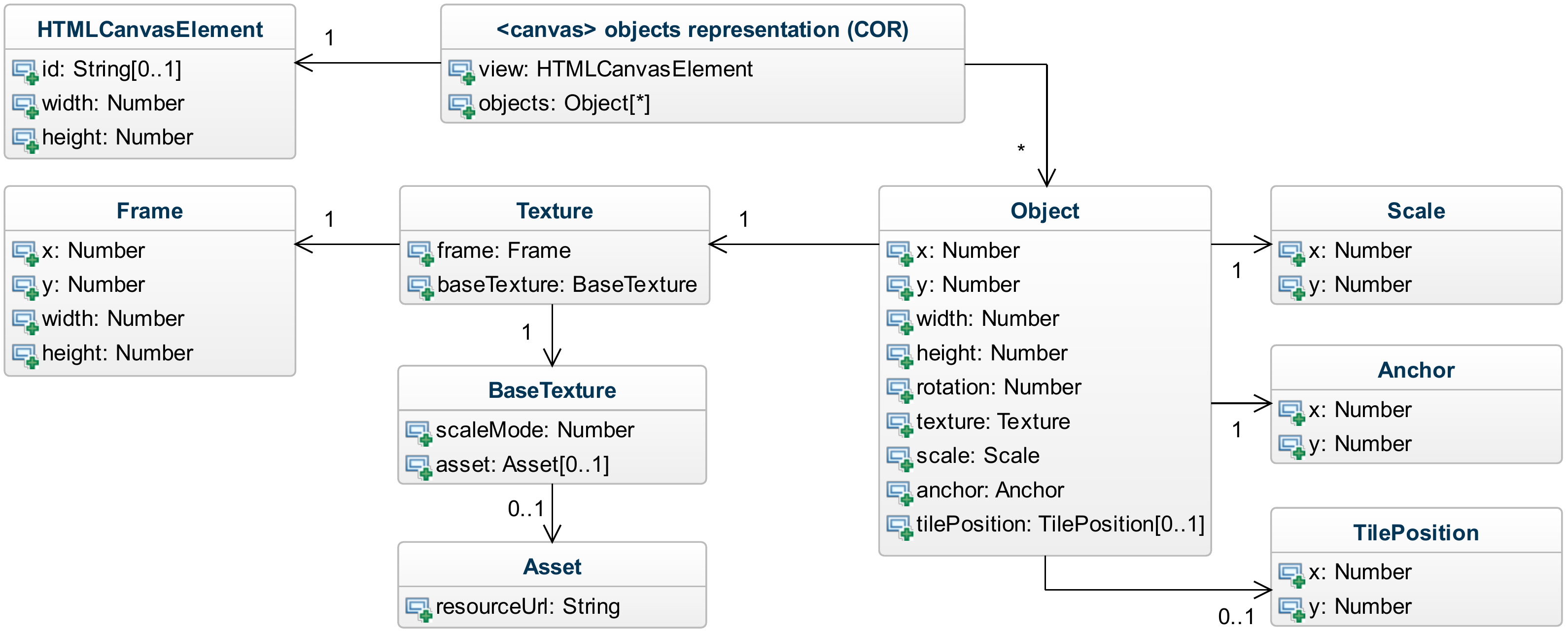}
    \caption{Unified modeling language (UML) class diagram for a \canvas objects representation (COR).}
    \label{fig:database}
\end{figure*}

For each snapshot, we automatically collect a screenshot and its respective COR. 
Figure~\ref{fig:database} illustrates what a COR contains in our approach.
A COR is used by a \canvas game to determine how to render game objects to the \canvas, such as the player character, background layers, and projectiles.
Each object in the COR has properties such as position, size, and rotation.

While performing a snapshot, we prevent new animation frames from being rendered, and save a frozen copy of the COR along with a synchronized screenshot of the current animation frame (as rendered to the \canvas).
Although our snapshot operation briefly prevents the rendering of a few new frames, it does not necessarily interrupt the main game loop (depending upon how a game is implemented).

As described in Section~\ref{subsec:canvasgames} of our paper, assets in \canvas games are served through web requests, and so we created a custom crawler to collect assets based on the resource URLs of objects in a \canvas game.
As can be seen in Figure~\ref{fig:database}, game objects are linked to their respective assets in the COR, meaning associating a game object with its respective asset is straightforward.

\subsection{Preprocessing images}\label{subsec:preprocessingassets}
\begin{figure*}[!t]
	\centering
	\includegraphics[width=\linewidth]{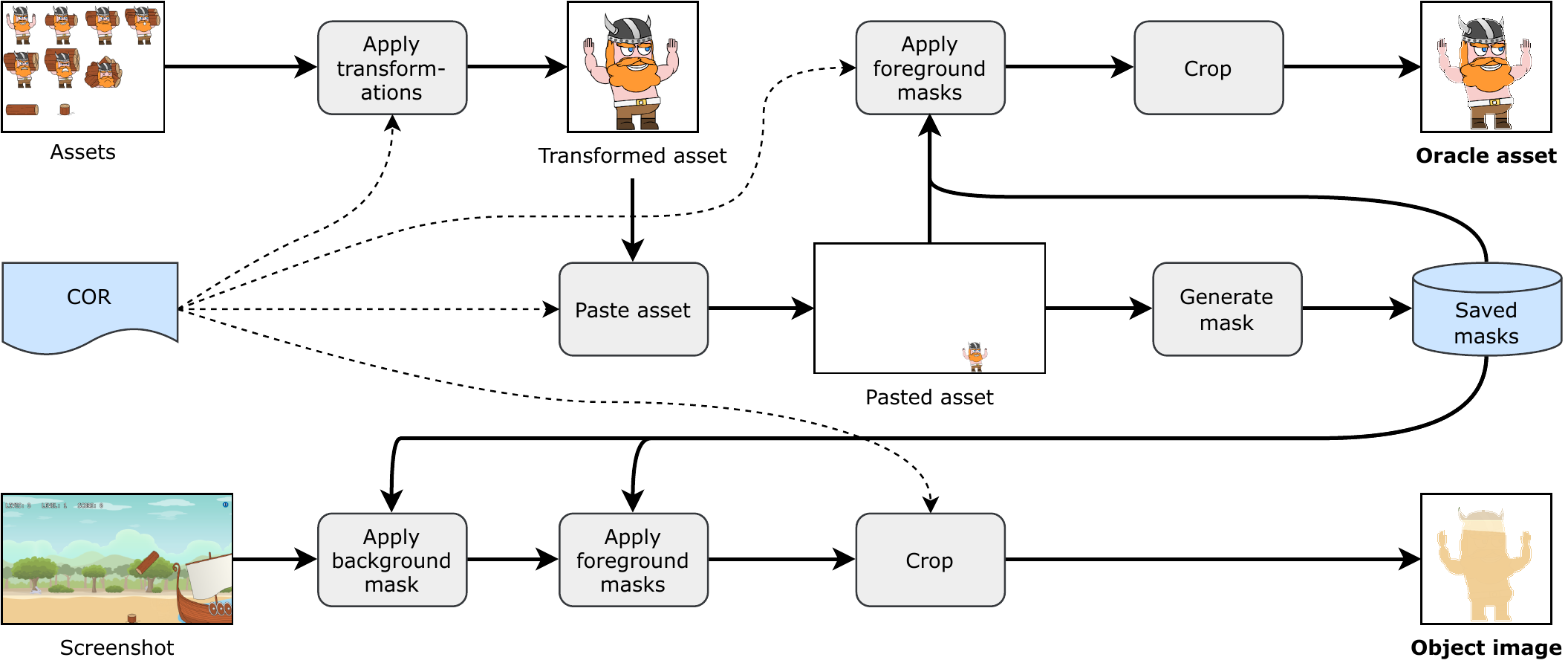}
	\caption{Automated image preprocessing pipeline (shown with visual bug \hyperref[tab:bugdetection]{\texttt{S1}}, \emph{player-character is invisible}).}
	\label{fig:preprocessing}
\end{figure*}

For each snapshot, we leverage the COR to automatically generate oracle assets and extract object images for comparison.
Figure~\ref{fig:preprocessing} shows our automated image processing pipeline.
Below, we detail our preprocessing steps for oracle assets and object images.

\begin{figure*}[!t]
	\setlength{\fboxsep}{0pt}%
	\centering
	\hfill
	\begin{subfigure}[t]{0.24\linewidth}
		\centering
		\fbox{\includegraphics[width=\textwidth]{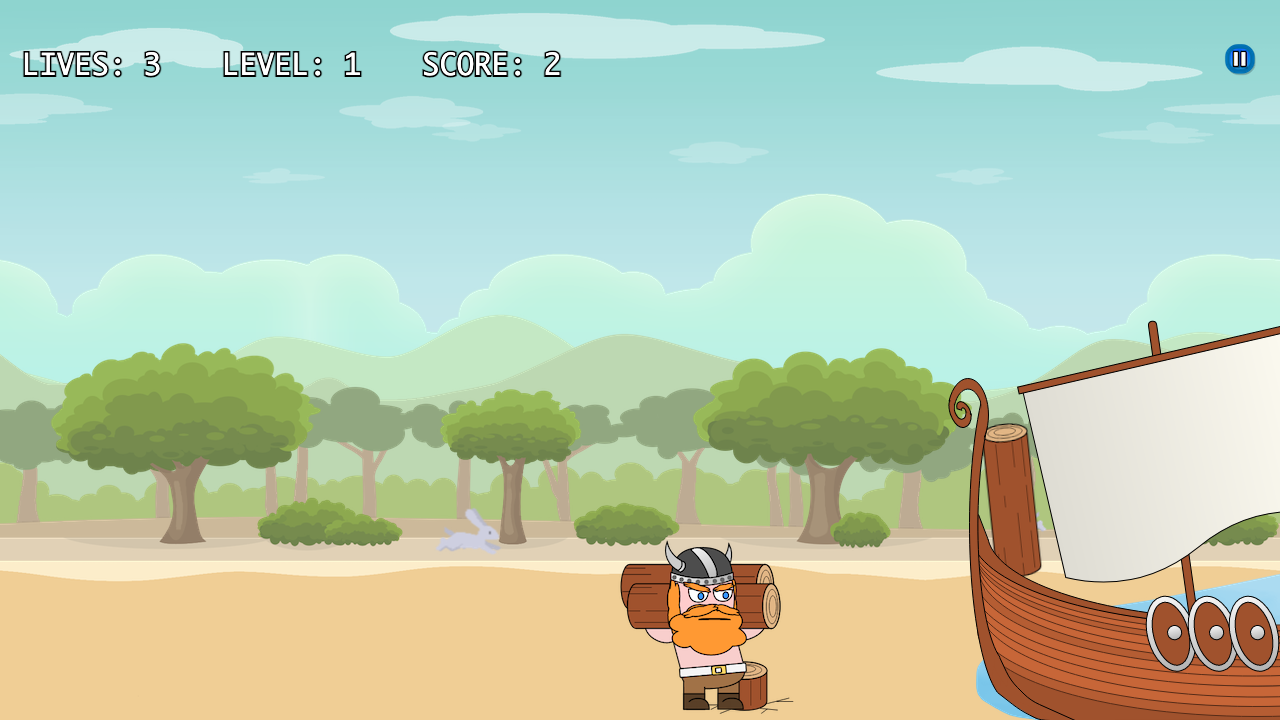}}
		\caption{Input screenshot}
		\label{fig:mask_input}
	\end{subfigure}
	\hfill
	\begin{subfigure}[t]{0.24\linewidth}
		\centering
		\fbox{\includegraphics[width=\textwidth]{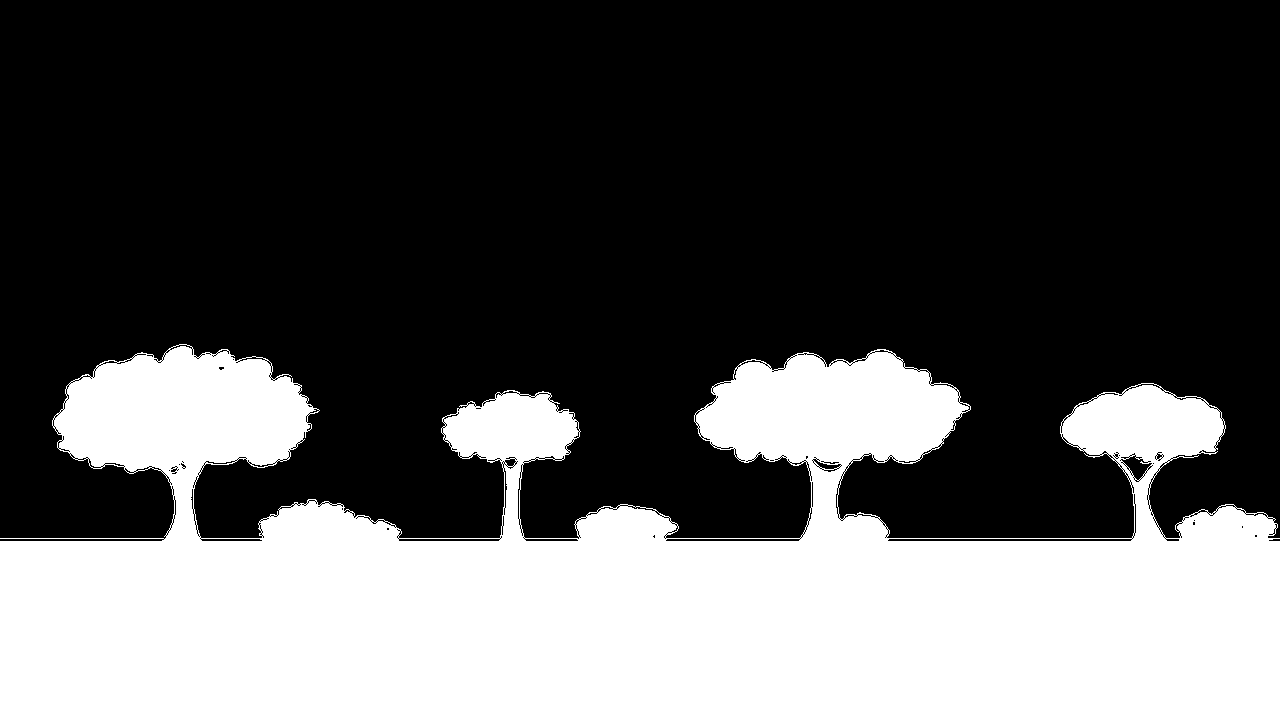}}
		\caption{The background mask}
		\label{fig:mask_bg}
	\end{subfigure}
	\hfill
	\begin{subfigure}[t]{0.24\linewidth}
		\centering
		\fbox{\includegraphics[width=\textwidth]{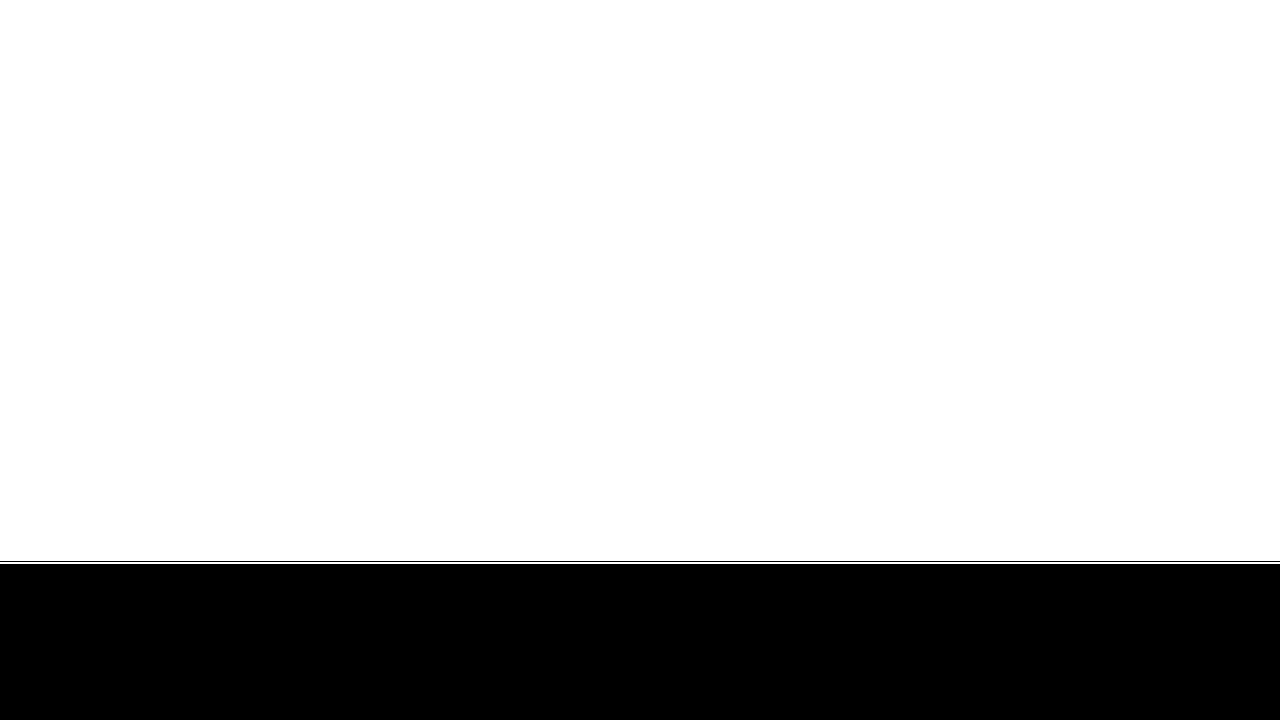}}
		\caption{One of the foreground masks}
		\label{fig:mask_fg}
	\end{subfigure}
	\hfill
	\begin{subfigure}[t]{0.24\linewidth}
		\centering
		\fbox{\includegraphics[width=\textwidth]{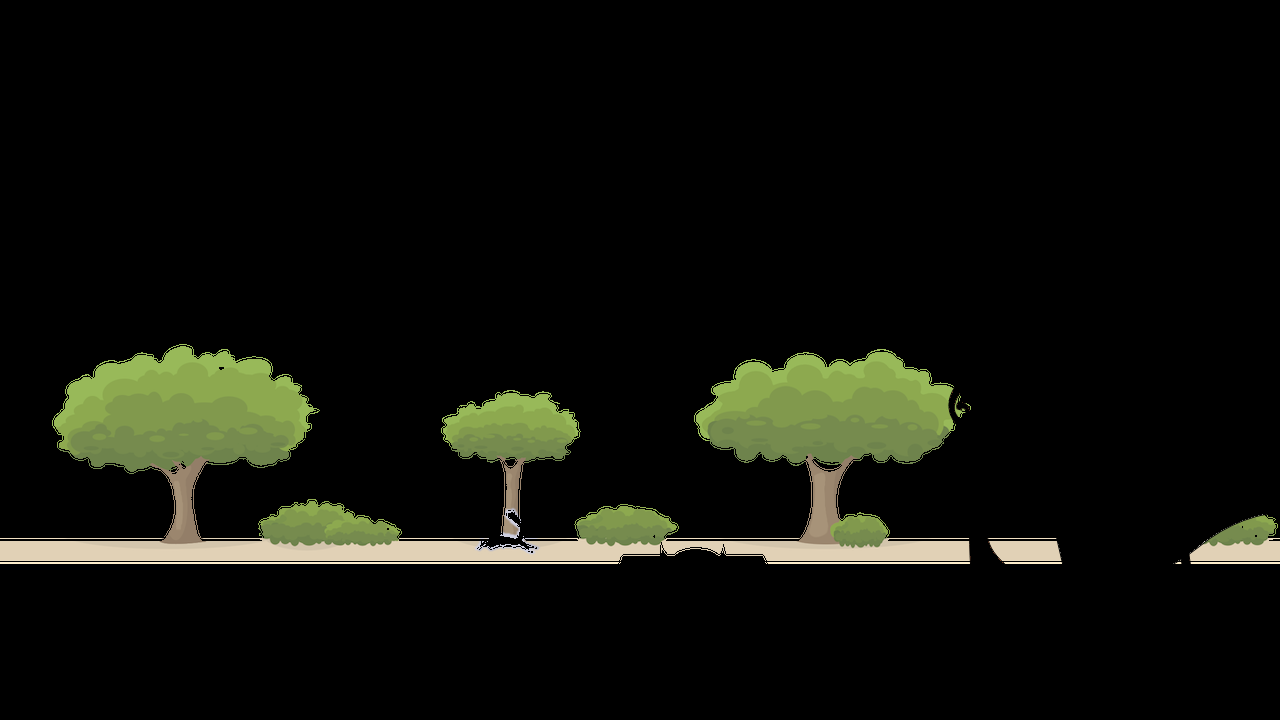}}
		\caption{Output object image}
		\label{fig:mask_output}
	\end{subfigure}
	\hfill
	\caption{A background mask and all overlapping foreground masks are applied to isolate object images in our approach.}
	\label{fig:masking}
\end{figure*}

We automatically preprocess game assets to generate oracle assets during the execution of a test using the following process:
\begin{enumerate}
    \item Apply transformations to the asset as specified in the COR. For example, crop, scale, tile, and/or rotate the asset.
    \item Paste the asset onto a blank image that is the same size as the \canvas. The paste location is determined by the COR, and will match the location of the game object in the screenshot if no bugs are present.
    \item Generate an image mask from the pasted asset (i.e., the result from the previous step) and save for later masking operations.
    \item For any overlapping objects, apply their saved masks over top of the pasted asset. Figure~\ref{fig:masking} shows an example of what this might look like.
    \item Crop the pasted asset.
\end{enumerate} 

We automatically decompose screenshots into a set of individual object images according to the following process:
\begin{enumerate}
    \item Apply the background mask, i.e., the mask generated from the object's respective asset.
    \item Apply the foreground masks, i.e., the masks that were generated from assets belonging to overlapping objects.
    \item Crop the object image out of the screenshot.
\end{enumerate}
After preprocessing, we have a set of image pairs, with each pair containing an oracle asset and object image, which should be exactly the same if no visual bugs are present.

\subsection{Detecting visual bugs}
For each pair of oracle asset and object image, we use an image similarity metric to automatically perform a visual comparison of the images.
Our approach relies on a threshold for this similarity metric to decide if a visual test case should pass or fail. 
This threshold should be defined empirically and for each game, as different games may have different levels of in-game randomness that could affect the similarity metric.
\section{Experiment Setup}\label{sec:experiments}
In this section, we describe our experiment setup for evaluating the performance of our approach and the baseline approach for automatically detecting visual bugs.
We selected snapshot testing as the baseline approach for comparison with our approach, as snapshot testing is the industry standard approach for detecting visual bugs in \canvas applications.

\subsection{The test game}
\begin{figure*}[t]
    \centering
    \includegraphics[width=\linewidth]{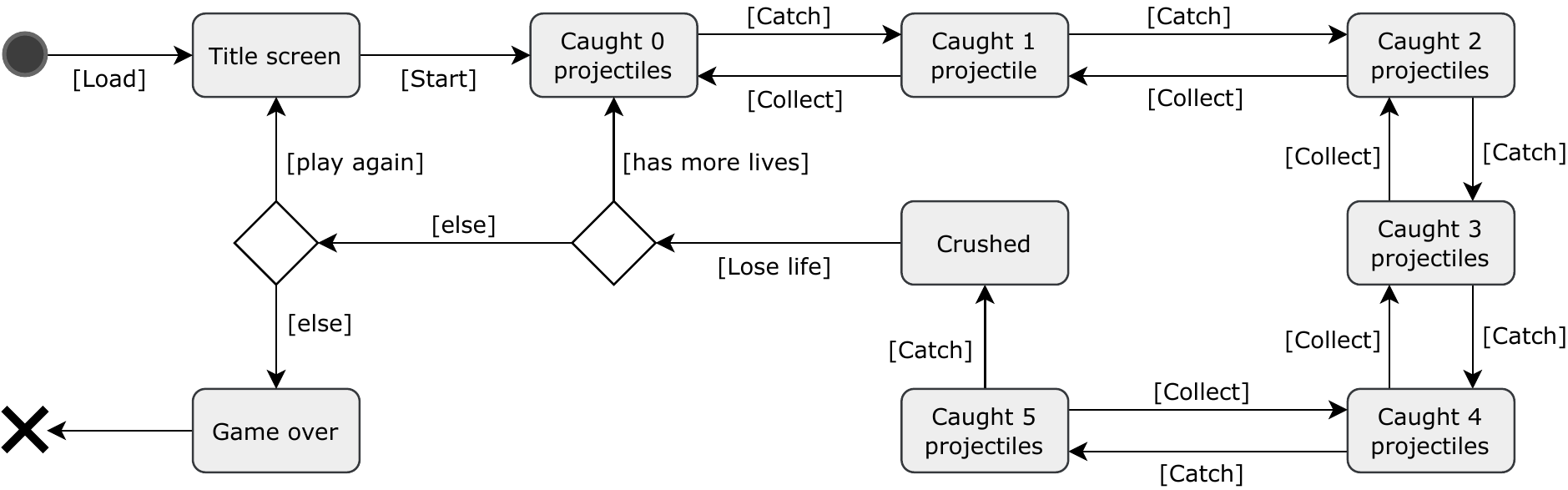}
    \caption{Finite state machine diagram for our test \canvas game.}
    \label{fig:testgame}
\end{figure*}

To evaluate our approach, we created a custom \canvas game using the \texttt{PixiJS}\footnote{\href{https://pixijs.com/}{https://pixijs.com/}} library and a freely available asset pack\footnote{\href{https://raventale.itch.io/parallax-background}{https://raventale.itch.io/parallax-background}}.
\texttt{PixiJS} is a popular free and open-source \canvas rendering library for creating 2D animations with the \canvas.
Figure~\ref{fig:testgame} shows the state machine diagram for our test game.
Our test game is a so-called catching game, i.e., a game in which projectiles are randomly thrown for the player to catch.
The test game contains a variety of animations, including animated sprites, rotating sprites, and background tiling sprites.
The test game was designed to be played at a resolution of 720p, with a maximum frame rate of $60$ FPS.

\subsection{The test case}
We wrote an automated test case for our \canvas game.
In our test case, the game was automatically opened in a browser window with size $1280px\times720px$.
Next, the game was started through an automated user click, and then the player-character was moved back and forth across the screen with automated mouse movements until the player lost a life (after which, the test case ended).
During each test case execution, 10 snapshots were taken.

\subsection{Injected visual bugs} \label{subsec:injectedbugs}
We evaluated the performance of the approaches by injecting visual bugs into our test game.
To target bugs that are relevant to \canvas games, we used the taxonomy of \canvas bugs constructed by Macklon et al.~\cite{macklon2022taxonomy}, and verified with an industrial partner that our injected bugs were relevant to industrial \canvas games.
In Table~\ref{tab:visualbugtypes}, we provide each visual bug type and an example description of a bug of that type as defined in the taxonomy of \canvas bugs.

\begin{figure}[!t]
	\centering
	\setlength{\fboxsep}{0pt}%
	\begin{subfigure}[t]{0.46\linewidth}
		\centering
		\fbox{\includegraphics[width=\textwidth]{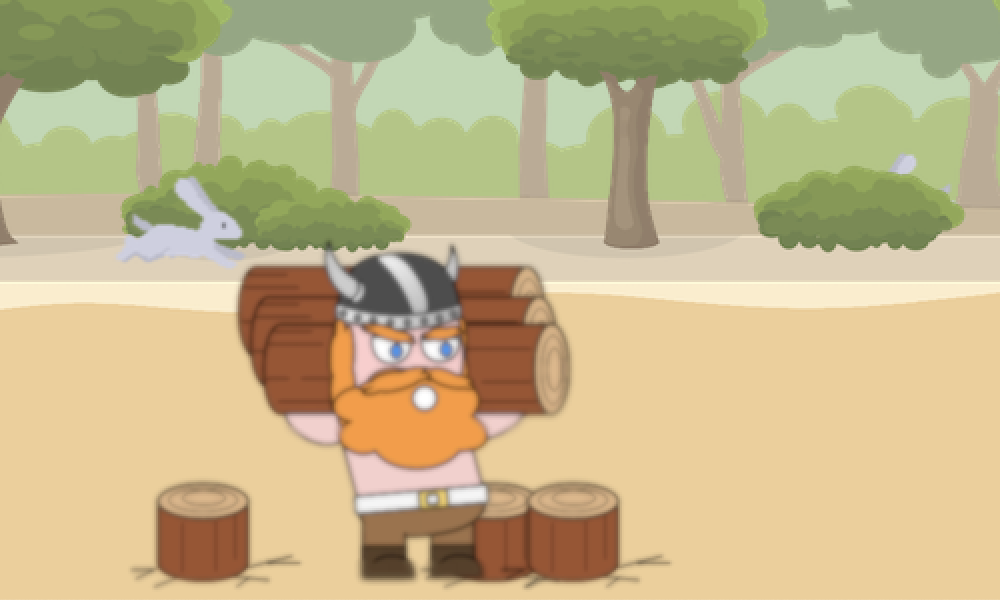}}
		\caption{Rendering bug \hyperref[tab:bugdetection]{\texttt{R3}} in Table~\ref{tab:bugdetection}. \textit{Viking and logs are blurred.}}
		\label{fig:rendering}
	\end{subfigure}
	\hfill
	\begin{subfigure}[t]{0.46\linewidth}
		\centering
		\fbox{\includegraphics[width=\textwidth]{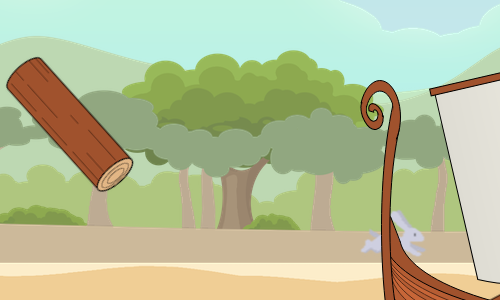}}
		\caption{Layout bug \hyperref[tab:bugdetection]{\texttt{L5}} in Table~\ref{tab:bugdetection}. \textit{Trees are in the wrong layer.}}
		\label{fig:layout}
	\end{subfigure}
	\hfill
	\begin{subfigure}[t]{0.46\linewidth}
		\centering
		\fbox{\includegraphics[width=\textwidth]{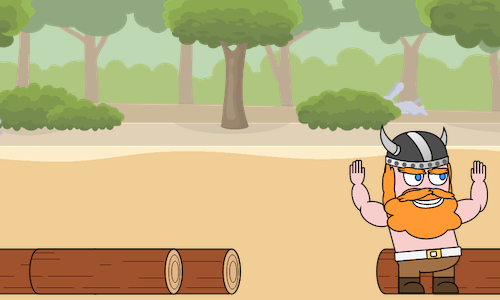}}
		\caption{State bug \hyperref[tab:bugdetection]{\texttt{S5}} in Table~\ref{tab:bugdetection}. \textit{Fallen log animation is not updating.}}
		\label{fig:state}
	\end{subfigure}
	\hfill
	\begin{subfigure}[t]{0.46\linewidth}
		\centering
		\fbox{\includegraphics[width=\textwidth]{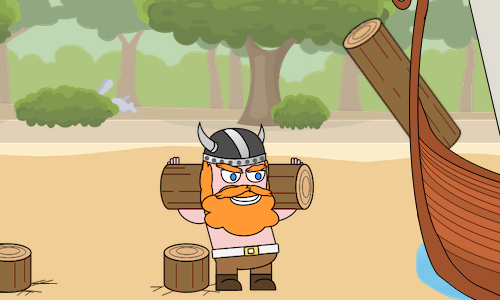}}
		\caption{Appearance bug \hyperref[tab:bugdetection]{\texttt{A4}}  in Table~\ref{tab:bugdetection}. \textit{Logs are a different colour.}}
		\label{fig:appearance}
	\end{subfigure}
	\caption{Sample instances of our injected visual bugs.}
	\label{fig:bug_samples}
\end{figure}

For each of the four visual bug types defined in the taxonomy of \canvas bugs, we injected six different bugs, with some primarily affecting foreground objects, and others primarily affecting background objects. 
In total, we injected 24 visual bugs.  Figure~\ref{fig:bug_samples} shows four example instances of visual bugs we injected into the test game, while Table~\ref{tab:bugdetection} provides detailed descriptions of each injected bug.

We injected most of the visual bugs by altering an asset during test execution, and then replaced it with the non-bugged (original) asset at the preprocessing stage of our approach.
We injected most of the visual bugs this way because real visual bugs can be very complex and difficult to reproduce~\cite{macklon2022taxonomy}.
Although our injected visual bugs had a different root cause than real visual bugs on the \canvas, we confirmed with an industrial partner that the visual effects were similar to visual bugs found in real \canvas games, meaning that our injected visual bugs were suitable for evaluating our approach.

\begin{table}[t]
    \centering
    \caption{Visual bug types found in \canvas applications~\cite{macklon2022taxonomy}.}
    \begin{tabular*}{\linewidth}{l@{\extracolsep{\fill}}l}
    \toprule
        \textbf{Type} & \textbf{Example Description} \\
    \midrule
        State &  Object visible but should be invisible.\\
        Appearance & Object has incorrect colour.\\
        Layout & Object has incorrect position, size, layer, etc.\\
        Rendering & Object is distorted, blurry, or contains artifacts.\\
    \bottomrule
    \end{tabular*}
    \label{tab:visualbugtypes}
\end{table}

\subsection{Similarity metrics} \label{subsec:similaritymetrics}
In our experiments we compared images using four similarity metrics: percentage overlap, mean squared error, structural similarity, and embedding similarity.

\subsubsection*{Percentage overlap (PCT)}
We selected percentage overlap as a similarity metric because it is the simplest method for calculating the similarity of two images, and is used in industry-standard tools such as \texttt{Percy}.
We calculated the PCT for a pair of images by calculating the percentage of pixels that exactly match between the two images.

\subsubsection*{Mean squared error (MSE)}
We selected mean squared error as a similarity metric because it is widely used in image processing as an image quality index~\cite{wang2002universal, eskicioglu1995image}, i.e., to measure degradation between original and reconstructed images.
The MSE actually captures the amount of difference (i.e., lower is better) instead of similarity (i.e., higher is better).
We calculated the mean squared error for a pair of images using the  \texttt{scikit-image}\footnote{\href{https://scikit-image.org/}{https://scikit-image.org/}} library.

\subsubsection*{Structural similarity (SSIM)}
We selected structural similarity as a similarity metric because it is intended to be complementary to mean squared error as an image quality index~\cite{wang2004image}.
We used the \texttt{scikit-image} library to calculate structural similarity for each pair of images.

\subsubsection*{Embedding similarity (ESIM)}
Our fourth and final metric, embedding similarity, was the similarity of two images when represented as embeddings of an image classification model, i.e., a deep learning vision model.
Embeddings are the vision model's inner-layer representation(s) of an image, i.e., the feature representations of the image before the classification layer.

We implemented embedding similarity in our experiments by encoding the images as the embeddings of the final convolutional layer of the \texttt{ResNet-50} model pre-trained on the ImageNet dataset~\cite{he2015deep}.
These image embeddings had a feature volume of $(2048,7,7)$.
We selected the pre-trained \texttt{ResNet-50} model as the \texttt{ResNet} architecture is widely used for transfer learning applications~\cite{rezende2017malicious, chen2020unblind, luo2019making}.
We extracted the embeddings of the final convolutional layer of the \texttt{ResNet-50} model, as is done in prior work~\cite{taesiri2020video, fazli2021under}.
To calculate the similarity of the embeddings, we selected cosine similarity, a widely used similarity metric~\cite{zhang2018unreasonable, viggiato2021identifying, viggiatousing, taesiri2022clip}.

We used the pre-trained model in inference mode, meaning we did not have to perform any data labelling, training, or fine-tuning, i.e., we used the model out-of-the-box.
We performed inference with the pre-trained \texttt{ResNet-50} model on an NVIDIA Titan RTX graphics card.
We loaded the model from the \texttt{torchvision}\footnote{\href{https://pytorch.org/vision/stable/index.html}{https://pytorch.org/vision/stable/index.html}} library and used the \texttt{PyTorch}\footnote{\href{https://pytorch.org/}{https://pytorch.org/}} library to calculate cosine similarity.

\subsection{Empirical threshold selection}\label{subsubsec:thresholds}
We empirically selected a single threshold for each similarity metric used in each approach to decide whether a test case is buggy.
To empirically determine the thresholds, we calculated the similarities of all image pairs for 10 repetitions of test data with no bugs injected (i.e., with non-buggy snapshots), and took the overall lowest (or highest for MSE) similarity score for each metric as our thresholds. Hence, we chose the thresholds to yield zero false positives, as false positives result in a wasted effort from the game developer's perspective (as they need to investigate the false positive).

\subsection{Evaluating the experiments}
Here we describe the methods we used to evaluate our experiments.

\subsubsection*{Statistical significance and effect sizes}
We used the Mann-Whitney U test~\cite{mann1947test} to determine if the populations of similarity scores were statistically significantly different.
The Mann-Whitney U test is a non-parametric test that compares two distributions of unrelated populations to determine how much the populations statistically overlap, with some probability $p$. 
Generally, a $p$ value of less than $0.05$ indicates that the populations display a statistically significant difference, as a very low $p$ value indicates it is very unlikely that two populations are statistically similar.

To better understand the results of the Mann-Whitney U test, we also calculated Cliff's delta~\cite{cliff1993dominance} to determine the extent to which the populations of buggy and non-buggy similarity scores were different per metric.
To interpret the Cliff's delta values ($d$), we used the thresholds provided by Romano et al.~\cite{romano2006exploring} to determine the effect sizes, as done in prior work~\cite{kamienski2021empirical}.
The thresholds used were as follows:

\begin{math}
    \text{Effect size} =
    \begin{cases}
       \text{negligible} &\quad\text{if}\quad |d|\leq0.147 \\ 
       \text{small} &\quad\text{if}\quad 0.147 < |d| \leq 0.33 \\
       \text{medium} &\quad\text{if}\quad 0.33 < |d| \leq 0.474 \\
       \text{large} &\quad\text{if}\quad 0.474 < |d| \leq 1\\
     \end{cases}
\end{math}



\subsubsection*{Accuracy}
Our choice of threshold selection (Section~\ref{subsubsec:thresholds}) meant that it was only possible for there to be true positive (TP) and false negative (FN) cases in our results for visual bug detection.
Therefore, the best choice of evaluation metric was accuracy, which was calculated as follows: 
$\mathit{accuracy}\text{\space}=\frac{(\#\text{\space}\mathit{true}\text{\space}\mathit{positives})}{(\#\text{\space}\mathit{true}\text{\space}\mathit{positives})\text{\space}+\text{\space}(\#\text{\space}\mathit{false}\text{\space}\mathit{negatives})}$
\section{Results}\label{sec:results}
In this section, we present our experimental results for automatically detecting visual bugs with our approach and the baseline approach.
When using MSE, SSIM, or ESIM as the similarity metric, we find that our approach achieves an accuracy of $100\%$ for our 24 injected visual bugs, compared to an accuracy of $44.6\%$ with the baseline approach (with PCT as the similarity metric).
Our results show that our approach is much more effective for automatically detecting visual bugs in \canvas-based applications than the baseline approach (traditional snapshot testing).

\begin{figure*}[t]
    \centering
    \includegraphics[width=\linewidth]{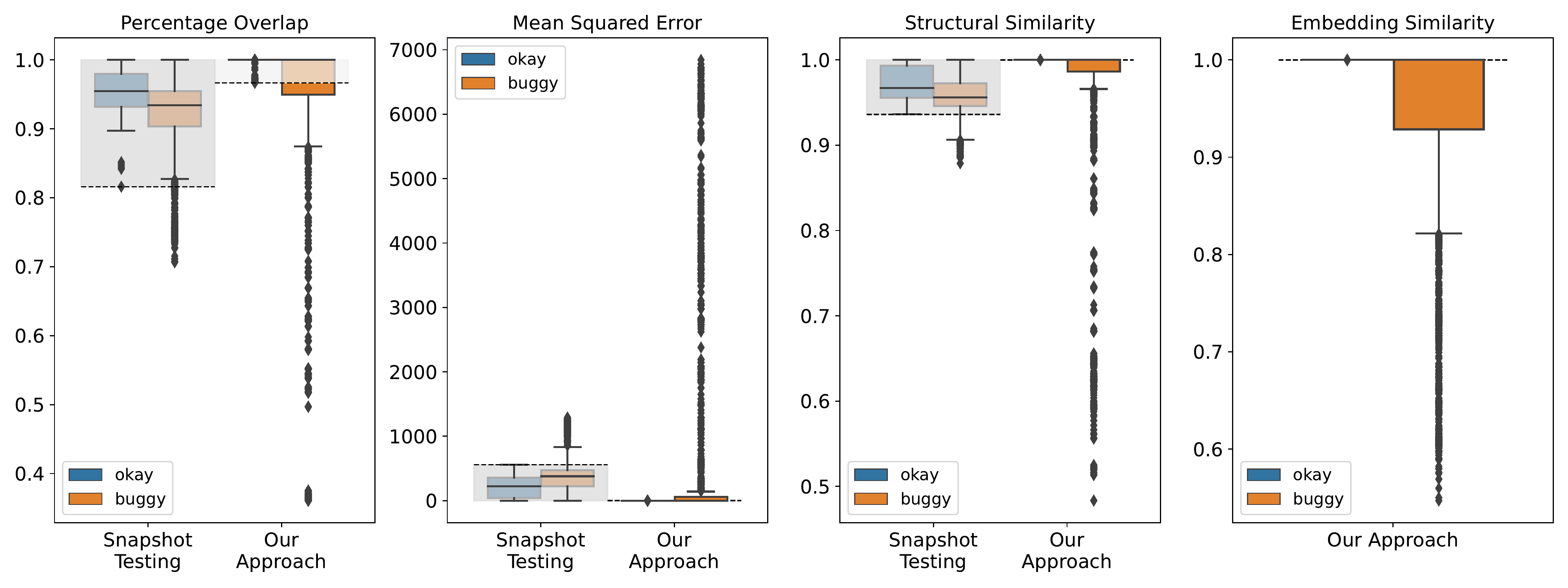}
    \caption{Boxplots of similarity scores for each evaluated similarity metric in each approach. Our selected thresholds are indicated by the grey dotted lines. Similarity scores in the greyed-out ranges are classified as non-buggy by each approach.}
    \label{fig:boxplots}
\end{figure*}

\subsection{Similarity scores}
Figure~\ref{fig:boxplots} shows the distributions of the similarity scores for each of the evaluated similarity metrics, with the minimum similarity for normal snapshots providing the thresholds for bug detection, as described in Section~\ref{subsubsec:thresholds}.
For each distribution, scores above the set threshold (within the greyed-out areas) would be accepted as within the normal range, whereas scores below the threshold would indicate a visual bug is present.
While the distributions are significantly different for all similarity metrics, the effect sizes show that there is a lot of overlap between the metrics when using the snapshot testing approach. 
As a result, a threshold is much harder to select for snapshot testing, and there will always be a trade-off between precision and recall. 
For our approach, the effect size (Table~\ref{tab:mwucliff}) is much larger indicating there is far less overlap between the distributions, allowing us to choose better-performing thresholds (which can also be observed from Figure~\ref{fig:boxplots}).
Clearly, there is much more overlap between normal and buggy cases when using snapshot testing than when using our approach with any of the similarity metrics.

\subsection{Bug detection}
Table~\ref{tab:bugdetection} shows the results for bug detection with each evaluated approach and similarity metric.
Our approach achieves a considerably higher accuracy (with any of our four evaluated similarity metrics) than snapshot testing.
In particular, our approach shows exciting potential for detecting visual bugs when MSE, SSIM, or ESIM is used as the similarity metric.
Our results indicate that only a single similarity metric is needed for our approach -- combining several metrics does not improve our overall results.

\begin{table}[t]
    \centering
    \caption{Mann-Whitney U test and Cliff's delta results.}
    \begin{tabular}{@{\extracolsep{\fill}}llc@{\extracolsep{\fill}}c}
    \toprule
     &  & \multicolumn{1}{c}{\textbf{Mann-Whitney U test}} & \multicolumn{1}{c}{\textbf{Cliff's delta}} \\
     & \textbf{Metric} &  \multicolumn{1}{c}{\textbf{Significant difference}} & \multicolumn{1}{c}{\textbf{Effect size}} \\
     \midrule
    \textbf{Snapshot} & \textbf{PCT} & \textit{yes} & \textit{small} \\
     \textbf{Testing} & \textbf{MSE} & \textit{yes} & \textit{medium} \\
     & \textbf{SSIM} & \textit{yes}& \textit{small} \\
     \midrule
    \textbf{Our} & \textbf{PCT} & \textit{yes}& \textit{medium} \\
    \textbf{Approach} & \textbf{MSE} & \textit{yes} & \textit{medium} \\
     & \textbf{SSIM} & \textit{yes}& \textit{medium} \\
     & \textbf{ESIM} & \textit{yes}& \textit{large} \\
     \bottomrule
    \end{tabular}
    \label{tab:mwucliff}
\end{table}

\begin{table*}[t] 
\centering
 \caption{Number of repetitions (out of 10) each visual bug was detected for each approach with each similarity metric. As detailed in Section~\ref{subsec:similaritymetrics}, we use the following similarity metrics: percentage overlap (PCT), mean squared error (MSE), structural similarity (SSIM), and embedding similarity (ESIM).}
 \label{tab:bugdetection}
\begin{tabular*}{\linewidth}{@{\extracolsep{\fill}}l@{\extracolsep{\fill}}l@{\extracolsep{\fill}}lrrrrrrr}
\toprule
&  &  & \multicolumn{3}{l}{\textbf{Snapshot Testing}} & \multicolumn{4}{l}{\textbf{Our Approach}} \\
\textbf{Type}                                                   & \textbf{Key} & \textbf{Bug Description}                       & \multicolumn{1}{l}{\textbf{PCT}}       & \multicolumn{1}{l}{\textbf{MSE}}       & \multicolumn{1}{l}{\textbf{SSIM}}      & \multicolumn{1}{l}{\textbf{PCT}}       & \multicolumn{1}{l}{\textbf{MSE}}        & \multicolumn{1}{l}{\textbf{SSIM}}       & \multicolumn{1}{l}{\textbf{ESIM}}       \\ \midrule
\multicolumn{1}{l}{\multirow{6}{*}{\rotatebox{90}{State}}}      & \texttt{S1}  & \textit{Viking is invisible.}                      & \cellcolor[HTML]{EFF9F4}1              & \cellcolor[HTML]{FFFFFF}0              & \cellcolor[HTML]{FFFFFF}0              & \cellcolor[HTML]{57BB8A}10             & \cellcolor[HTML]{57BB8A}10              & \cellcolor[HTML]{57BB8A}10              & \cellcolor[HTML]{57BB8A}10              \\
                                                                & \texttt{S2}  & \textit{A background hill is invisible.}       & \cellcolor[HTML]{79C9A2}8              & \cellcolor[HTML]{BCE4D1}4              & \cellcolor[HTML]{9BD7B9}6              & \cellcolor[HTML]{FFFFFF}0              & \cellcolor[HTML]{57BB8A}10              & \cellcolor[HTML]{57BB8A}10              & \cellcolor[HTML]{57BB8A}10              \\
                                                                & \texttt{S3}  & \textit{Ship is invisible.}                    & \cellcolor[HTML]{57BB8A}10             & \cellcolor[HTML]{57BB8A}10             & \cellcolor[HTML]{57BB8A}10             & \cellcolor[HTML]{57BB8A}10             & \cellcolor[HTML]{57BB8A}10              & \cellcolor[HTML]{57BB8A}10              & \cellcolor[HTML]{57BB8A}10              \\
                                                                & \texttt{S4}  & \textit{Viking animation is not updating.}            & \cellcolor[HTML]{EFF9F4}1              & \cellcolor[HTML]{FFFFFF}0              & \cellcolor[HTML]{DEF2E8}2              & \cellcolor[HTML]{57BB8A}10             & \cellcolor[HTML]{57BB8A}10              & \cellcolor[HTML]{57BB8A}10              & \cellcolor[HTML]{57BB8A}10              \\
                                                                & \texttt{S5}  & \textit{Fallen log animation is not updating.}    & \cellcolor[HTML]{8AD0AE}7              & \cellcolor[HTML]{CDEBDC}3              & \cellcolor[HTML]{ABDDC5}5              & \cellcolor[HTML]{57BB8A}10             & \cellcolor[HTML]{57BB8A}10              & \cellcolor[HTML]{57BB8A}10              & \cellcolor[HTML]{57BB8A}10              \\
                                                                & \texttt{S6}  & \textit{Button should be hidden.}              & \cellcolor[HTML]{ABDDC5}5              & \cellcolor[HTML]{57BB8A}10             & \cellcolor[HTML]{79C9A2}8              & \cellcolor[HTML]{FFFFFF}0              & \cellcolor[HTML]{57BB8A}10              & \cellcolor[HTML]{57BB8A}10              & \cellcolor[HTML]{57BB8A}10              \\
\midrule
\multicolumn{1}{l}{\multirow{6}{*}{\rotatebox{90}{Appearance}}} & \texttt{A1}  & \textit{Viking has the wrong beard colour.}        & \cellcolor[HTML]{EFF9F4}1              & \cellcolor[HTML]{BCE4D1}4              & \cellcolor[HTML]{BCE4D1}4              & \cellcolor[HTML]{57BB8A}10             & \cellcolor[HTML]{57BB8A}10              & \cellcolor[HTML]{57BB8A}10              & \cellcolor[HTML]{57BB8A}10              \\
                                                                & \texttt{A2}  & \textit{Entire viking has the wrong colour.}        & \cellcolor[HTML]{DEF2E8}2              & \cellcolor[HTML]{8AD0AE}7              & \cellcolor[HTML]{CDEBDC}3              & \cellcolor[HTML]{57BB8A}10             & \cellcolor[HTML]{57BB8A}10              & \cellcolor[HTML]{57BB8A}10              & \cellcolor[HTML]{57BB8A}10              \\
                                                                & \texttt{A3}  & \textit{Viking and logs are grey-scaled.}          & \cellcolor[HTML]{CDEBDC}3              & \cellcolor[HTML]{FFFFFF}0              & \cellcolor[HTML]{CDEBDC}3              & \cellcolor[HTML]{57BB8A}10             & \cellcolor[HTML]{57BB8A}10              & \cellcolor[HTML]{57BB8A}10              & \cellcolor[HTML]{57BB8A}10              \\
                                                                & \texttt{A4}  & \textit{Logs have the wrong colour.}            & \cellcolor[HTML]{CDEBDC}3              & \cellcolor[HTML]{EFF9F4}1              & \cellcolor[HTML]{BCE4D1}4              & \cellcolor[HTML]{57BB8A}10             & \cellcolor[HTML]{57BB8A}10              & \cellcolor[HTML]{57BB8A}10              & \cellcolor[HTML]{57BB8A}10              \\
                                                                & \texttt{A5}  & \textit{The ship's sail has the wrong colour.}            & \cellcolor[HTML]{68C296}9              & \cellcolor[HTML]{CDEBDC}3              & \cellcolor[HTML]{8AD0AE}7              & \cellcolor[HTML]{57BB8A}10             & \cellcolor[HTML]{57BB8A}10              & \cellcolor[HTML]{57BB8A}10              & \cellcolor[HTML]{57BB8A}10              \\
                                                                & \texttt{A6}  & \textit{A background bunny has the wrong colour.}    & \cellcolor[HTML]{CDEBDC}3              & \cellcolor[HTML]{CDEBDC}3              & \cellcolor[HTML]{ABDDC5}5              & \cellcolor[HTML]{FFFFFF}0              & \cellcolor[HTML]{57BB8A}10              & \cellcolor[HTML]{57BB8A}10              & \cellcolor[HTML]{57BB8A}10              \\
\midrule
\multicolumn{1}{l}{\multirow{6}{*}{\rotatebox{90}{Layout}}}     & \texttt{L1}  & \textit{Ship is in the wrong location.}     & \cellcolor[HTML]{57BB8A}10             & \cellcolor[HTML]{57BB8A}10             & \cellcolor[HTML]{57BB8A}10             & \cellcolor[HTML]{57BB8A}10             & \cellcolor[HTML]{57BB8A}10              & \cellcolor[HTML]{57BB8A}10              & \cellcolor[HTML]{57BB8A}10              \\
                                                                & \texttt{L2}  & \textit{Viking is in the wrong location.}       & \cellcolor[HTML]{DEF2E8}2              & \cellcolor[HTML]{DEF2E8}2              & \cellcolor[HTML]{DEF2E8}2              & \cellcolor[HTML]{57BB8A}10             & \cellcolor[HTML]{57BB8A}10              & \cellcolor[HTML]{57BB8A}10              & \cellcolor[HTML]{57BB8A}10              \\
                                                                & \texttt{L3}  & \textit{Background clouds are in the wrong location.} & \cellcolor[HTML]{57BB8A}10             & \cellcolor[HTML]{ABDDC5}5              & \cellcolor[HTML]{CDEBDC}3              & \cellcolor[HTML]{57BB8A}10             & \cellcolor[HTML]{57BB8A}10              & \cellcolor[HTML]{57BB8A}10              & \cellcolor[HTML]{57BB8A}10              \\
                                                                & \texttt{L4}  & \textit{Viking has the wrong rotation.}                & \cellcolor[HTML]{DEF2E8}2              & \cellcolor[HTML]{FFFFFF}0              & \cellcolor[HTML]{DEF2E8}2              & \cellcolor[HTML]{57BB8A}10             & \cellcolor[HTML]{57BB8A}10              & \cellcolor[HTML]{57BB8A}10              & \cellcolor[HTML]{57BB8A}10              \\
                                                                & \texttt{L5}  & \textit{Background trees are in the wrong layer.}    & \cellcolor[HTML]{79C9A2}8              & \cellcolor[HTML]{8AD0AE}7              & \cellcolor[HTML]{57BB8A}10             & \cellcolor[HTML]{57BB8A}10             & \cellcolor[HTML]{57BB8A}10              & \cellcolor[HTML]{57BB8A}10              & \cellcolor[HTML]{57BB8A}10              \\
                                                                & \texttt{L6}  & \textit{Logs have the wrong rotation.}           & \cellcolor[HTML]{DEF2E8}2              & \cellcolor[HTML]{EFF9F4}1              & \cellcolor[HTML]{EFF9F4}1              & \cellcolor[HTML]{57BB8A}10             & \cellcolor[HTML]{57BB8A}10              & \cellcolor[HTML]{57BB8A}10              & \cellcolor[HTML]{57BB8A}10              \\
\midrule
\multicolumn{1}{l}{\multirow{6}{*}{\rotatebox{90}{Rendering}}}  & \texttt{R1}  & \textit{Viking and logs are very distorted.}       & \cellcolor[HTML]{DEF2E8}2              & \cellcolor[HTML]{EFF9F4}1              & \cellcolor[HTML]{DEF2E8}2              & \cellcolor[HTML]{57BB8A}10             & \cellcolor[HTML]{57BB8A}10              & \cellcolor[HTML]{57BB8A}10              & \cellcolor[HTML]{57BB8A}10              \\
                                                                & \texttt{R2}  & \textit{Viking and logs are slightly distorted.}   & \cellcolor[HTML]{DEF2E8}2              & \cellcolor[HTML]{EFF9F4}1              & \cellcolor[HTML]{CDEBDC}3              & \cellcolor[HTML]{57BB8A}10             & \cellcolor[HTML]{57BB8A}10              & \cellcolor[HTML]{57BB8A}10              & \cellcolor[HTML]{57BB8A}10              \\
                                                                & \texttt{R3}  & \textit{Viking and logs are blurred.}              & \cellcolor[HTML]{DEF2E8}2              & \cellcolor[HTML]{FFFFFF}0              & \cellcolor[HTML]{FFFFFF}0              & \cellcolor[HTML]{57BB8A}10             & \cellcolor[HTML]{57BB8A}10              & \cellcolor[HTML]{57BB8A}10              & \cellcolor[HTML]{57BB8A}10              \\
                                                                & \texttt{R4}  & \textit{Background trees covered in patches.}  & \cellcolor[HTML]{8AD0AE}7              & \cellcolor[HTML]{79C9A2}8              & \cellcolor[HTML]{79C9A2}8              & \cellcolor[HTML]{FFFFFF}0              & \cellcolor[HTML]{57BB8A}10              & \cellcolor[HTML]{57BB8A}10              & \cellcolor[HTML]{57BB8A}10              \\
                                                                & \texttt{R5}  & \textit{Background bushes have artifacts.}       & \cellcolor[HTML]{DEF2E8}2              & \cellcolor[HTML]{FFFFFF}0              & \cellcolor[HTML]{DEF2E8}2              & \cellcolor[HTML]{FFFFFF}0              & \cellcolor[HTML]{57BB8A}10              & \cellcolor[HTML]{57BB8A}10              & \cellcolor[HTML]{57BB8A}10              \\
                                                                & \texttt{R6}  & \textit{Background beach has tearing.}         & \cellcolor[HTML]{ABDDC5}5              & \cellcolor[HTML]{FFFFFF}0              & \cellcolor[HTML]{8AD0AE}7              & \cellcolor[HTML]{FFFFFF}0              & \cellcolor[HTML]{57BB8A}10              & \cellcolor[HTML]{57BB8A}10              & \cellcolor[HTML]{57BB8A}10              \\
\midrule
\textbf{Accuracy}                                               &              &                                                & \cellcolor[HTML]{B5E1CB}\textbf{44.6\%} & \cellcolor[HTML]{C7E9D8}\textbf{33.3\%} & \cellcolor[HTML]{B5E1CB}\textbf{44.6\%} & \cellcolor[HTML]{81CCA8}\textbf{75.0\%} & \cellcolor[HTML]{57BB8A}\textbf{100.0\%} & \cellcolor[HTML]{57BB8A}\textbf{100.0\%} & \cellcolor[HTML]{57BB8A}\textbf{100.0\%} \\ 
\bottomrule
\end{tabular*}
\end{table*}

\subsection{Execution duration}
To better understand the performance of our approach, we timed the executions of our approach and the baseline approach.
Our approach took considerably longer (3 additional seconds per snapshot) to run than the baseline approach. 
However, the accuracy of the baseline approach indicates that it is not a very useful one in practice.
The bulk of time in our approach is spent preprocessing the images, whereas calculating the similarities is relatively quick, with the exception of SSIM.
In practice, we would not have to compute SSIM, because MSE, SSIM, and ESIM provide similar accuracy in our experiments.

\section{Threats to Validity}\label{sec:threats}

\subsubsection*{Construct validity}
Our results may be biased towards the set of visual bugs that we injected in our experiments.
However, our injected visual bugs covered all four visual bug types that are relevant to the \canvas, as defined in prior work~\cite{macklon2022taxonomy}. 
While the visual bugs we injected may not have the same cause as real visual bugs found in \canvas applications, the visual effects are the same; each injected visual bug was designed to resemble a real world example.
To mitigate the threat of injecting unrealistic bugs we also confirmed with an industrial partner that our injected bugs were representative of real visual bugs in industrial \canvas games.

There are different possible choices of image comparison metrics for snapshot testing, the baseline approach used in our experiments.
We selected PCT as an image comparison metric because \texttt{Percy}, a widely used snapshot testing tool, uses a threshold-based image comparison metric that is similar to PCT. 
We also empirically evaluated MSE and SSIM for snapshot testing, and determined that neither were better than PCT for snapshot testing.

\subsubsection*{Internal validity}
In our approach, we utilized the \canvas objects representation (COR) combined with the game assets to automatically generate visual test oracles (i.e., oracle assets).
Our approach therefore assumes that no bugs originate in these parts of the game.
It is fair to assume that the assets provide accurate baselines for comparison with the rendered game objects on the \canvas.
In addition, it is fair to assume that the COR can be used to generate test oracles for detecting visual bugs, as any bug present in the COR would not be a visual bug.

A threat to internal validity is our choice of background fill colour when applying masks during image preprocessing in our experiments.
A fill colour must be selected to fill the blank space that results from the masking operations.
To address this threat, we ran our experiments with three different fill colours in 8-bit RGBA format: (0, 0, 0, 255), (255, 255, 255, 255), and (255, 0, 0, 255).
We empirically determined that for our test game, changing the fill colour for masking only affected our results with embedding similarity (ESIM), indicating that our use of ESIM may not be appropriate due to our extensive preprocessing.

A threat to internal validity is related to how we handle assets that have partial transparency in our experiments.
In our experiments, we chose to remove all partial transparency (i.e., make it fully transparent), as we empirically determined that this choice provided the best performance.
However, this means that we may miss some visual bugs that (primarily) affect the partially transparent areas of an object on the \canvas.
More work is required to better handle assets with partial transparency when generating masks.

\subsubsection*{External validity}
Our approach has only been evaluated with a single \canvas game.
Thresholds for detecting visual bugs with our approach will most likely differ on a per-game basis, and not all games may have as clear similarity thresholds as those found for our test game.
Therefore, automatically setting the thresholds to detect visual bugs may not be as effective for other \canvas games, meaning manual adjustment may be required.
We designed our test game with the randomness and a variety of animations that were inspired by the visual styles and effects of industrial \canvas games. 
However, future studies should investigate further how different styles of games impact the performance of our approach.

Our approach is evaluated only a 2D \canvas game that was built with the \texttt{PixiJS} \canvas rendering framework.
More work is required to understand how well our approach works for other 2D \canvas games, 3D \canvas games, and non-\canvas games.

In our experiments we leverage an existing \canvas objects representation (COR) provided by \texttt{PixiJS}.
If a COR is not available in a \canvas game, our approach would not work for that game.
Similarly, in our experiments we leverage existing \canvas game assets to generate visual test oracles during the test, but if a \canvas game does not use assets for its graphics rendering, then our approach would not work for that game.

Some types of graphics (e.g., skeletal animations, particle effects) are not in our test game, and are therefore not accounted for in the implementation of our approach.
More work is required to understand how our method performs when implemented for other types of graphics that are common in 2D \canvas games.
\section{Conclusion}\label{sec:conclusion}
In this paper, we present a novel approach for automatically detecting visual bugs in \canvas games.
By leveraging the \canvas objects representation (COR), we are able to automatically generate oracle assets for comparison with isolated object images (as rendered to the \canvas) and detect a wide variety of visual bugs in \canvas games.
We found that our approach far outperforms the current industry standard approach (traditional snapshot testing) for automatically detecting visual bugs in \canvas games.
We evaluated four similarity metrics with our approach, and found that mean squared error (MSE), structural similarity (SSIM), and embedding similarity (ESIM) each provided an accuracy of $100\%$ for our 24 injected visual bugs.
An implementation of our approach and our testbed is available at the following link: \href{https://github.com/asgaardlab/canvas-visual-bugs-testbed}{https://github.com/asgaardlab/canvas-visual-bugs-testbed}.

\section*{Acknowledgments}
The research reported in this article has been supported by Prodigy Education and the Natural Sciences and Engineering Research Council of Canada under the Alliance Grant project ALLRP 550309.

\balance
\printbibliography

@online{goodboy2020playco,
    author = {Playco},
    title = {Playco, Global Leader in Instant Games, Acquires {G}oodboy {D}igital, Creators of {P}ixi{JS}, the Number One {HTML5} Game Engine.},
    url = {https://www.play.co/press/playco-official-press-release-210928-en?utm_sq=guxn7m9l5r},
    year = {2021},
    month = {09},
    day = {28},
    note={Last accessed 5 May 2022.},
}

@online{google2020gamesnacks,
    author = {Ani Mohan},
    title = {{G}ame{S}nacks brings quick, casual games to any device},
    url = {https://blog.google/technology/area-120/gamesnacks-brings-quick-casual-games-any-device/},
    year = {2020},
    month = {02},
    day = {13},
    note={Last accessed 4 May 2022.},
}

@misc{html5gamedevs:0,
    url={https://www.html5gamedevs.com/forum/13-frameworks/},
    author={{HTML5} {G}ame {D}evs},
    title={Forum - Frameworks},
    note={Last accessed 4 May 2022.},
    year={2022}
}

@inproceedings{almansoury2020investigating,
  title={{I}nvestigating {W}eb{3D} topics on {S}tack{O}verflow: a preliminary study of {W}eb{GL} and {T}hree.js},
  author={Almansoury, Farag and Kpodjedo, S{\`e}gla and Boussaidi, Ghizlane El},
  booktitle={The 25th International Conference on 3D Web Technology},
  pages={1--2},
  year={2020}
}

@book{parisi2012webgl,
  title={WebGL: up and running},
  author={Parisi, Tony},
  year={2012},
  publisher={O'Reilly Media, Inc.}
}

@book{parisi2014programming,
  title={Programming {3D} Applications with {HTML5} and {W}eb{GL}: {3D} Animation and Visualization for Web Pages},
  author={Parisi, Tony},
  year={2014},
  publisher={O'Reilly Media, Inc.}
}

@article{konstantinidis2016moving,
  title={Moving real exergaming engines on the web: the web{F}it{F}or{A}ll case study in an active and healthy ageing living lab environment},
  author={Konstantinidis, Evdokimos I and Bamparopoulos, Giorgos and Bamidis, Panagiotis D},
  journal={IEEE journal of biomedical and health informatics},
  volume={21},
  number={3},
  pages={859--866},
  year={2016},
  publisher={IEEE}
}

@inproceedings{yogya2014comparison,
  title={Comparison of physics frameworks for {W}eb{GL}-based game engine},
  author={Yogya, Resa and Kosala, Raymond},
  booktitle={EPJ Web of Conferences},
  volume={68},
  pages={00035},
  year={2014},
  organization={EDP Sciences}
}

@book{fulton2013html5,
  title={{HTML5} canvas: native interactivity and animation for the web},
  author={Fulton, Steve and Fulton, Jeff},
  year={2013},
  publisher={O'Reilly Media, Inc.}
}

@article{macklon2022taxonomy,
  title={{A} Taxonomy of {HTML5} Canvas Bugs},
  author={Macklon, Finlay and Viggiato, Markos and Romanova, Natalia and Buzon, Chris and Paas, Dale and Bezemer, Cor-Paul},
  journal={arXiv preprint arXiv:2201.07351},
  year={2022}
}

@inproceedings{bajammal2018web,
  title={Web canvas testing through visual inference},
  author={Bajammal, Mohammad and Mesbah, Ali},
  booktitle={2018 IEEE 11th International Conference on Software Testing, Verification and Validation (ICST)},
  pages={193--203},
  year={2018},
  organization={IEEE}
}

@inproceedings{issa2012visual,
  title={Visual testing of Graphical User Interfaces: An exploratory study towards systematic definitions and approaches},
  author={Issa, Ayman and Sillito, Jonathan and Garousi, Vahid},
  booktitle={2012 14th IEEE International Symposium on Web Systems Evolution (WSE)},
  pages={11--15},
  year={2012},
  organization={IEEE}
}

@article{bajammal2020survey,
  title={A Survey on the Use of Computer Vision to Improve Software Engineering Tasks},
  author={Bajammal, Mohammad and Stocco, Andrea and Mazinanian, Davood and Mesbah, Ali},
  journal={IEEE Transactions on Software Engineering},
  year={2020},
  publisher={IEEE}
}

@inproceedings{ricca2021ai,
  title={{AI}-based Test Automation: A Grey Literature Analysis},
  author={Ricca, Filippo and Marchetto, Alessandro and Stocco, Andrea},
  booktitle={2021 IEEE International Conference on Software Testing, Verification and Validation Workshops (ICSTW)},
  pages={263--270},
  year={2021},
  organization={IEEE}
}

@inproceedings{bajammal2021semantic,
  title={Semantic Web Accessibility Testing via Hierarchical Visual Analysis},
  author={Bajammal, Mohammad and Mesbah, Ali},
  booktitle={2021 IEEE/ACM 43rd International Conference on Software Engineering (ICSE)},
  pages={1610--1621},
  year={2021},
  organization={IEEE}
}

@article{bajammal2021page,
  title={Page Segmentation using Visual Adjacency Analysis},
  author={Bajammal, Mohammad and Mesbah, Ali},
  journal={arXiv preprint arXiv:2112.11975},
  year={2021}
}

@article{yandrapally2021fragment,
  title={Fragment-Based Test Generation For Web Apps},
  author={Yandrapally, Rahulkrishna and Mesbah, Ali},
  journal={arXiv preprint arXiv:2110.14043},
  year={2021}
}

@article{mazinanian2021style,
  title={Style-Guided Web Application Exploration},
  author={Mazinanian, Davood and Bajammal, Mohammad and Mesbah, Ali},
  journal={arXiv preprint arXiv:2111.12184},
  year={2021}
}

@inproceedings{white2019improving,
  title={Improving random GUI testing with image-based widget detection},
  author={White, Thomas D and Fraser, Gordon and Brown, Guy J},
  booktitle={Proceedings of the 28th ACM SIGSOFT International Symposium on Software Testing and Analysis},
  pages={307--317},
  year={2019}
}

@inproceedings{zhao2020seenomaly,
  title={Seenomaly: Vision-based linting of gui animation effects against design-don't guidelines},
  author={Zhao, Dehai and Xing, Zhenchang and Chen, Chunyang and Xu, Xiwei and Zhu, Liming and Li, Guoqiang and Wang, Jinshui},
  booktitle={2020 IEEE/ACM 42nd International Conference on Software Engineering (ICSE)},
  pages={1286--1297},
  year={2020},
  organization={IEEE}
}

@article{xue2022learning,
  title={Learning-Replay Based Automated Robotic Testing for Mobile App},
  author={Xue, Feng and Wu, Junsheng and Zhang, Tao},
  journal={Mobile Information Systems},
  volume={2022},
  year={2022},
  publisher={Hindawi}
}

@article{kamienski2021empirical,
  title={An empirical study of {Q\&A} websites for game developers},
  author={Kamienski, Arthur and Bezemer, Cor-Paul},
  journal={Empirical Software Engineering},
  volume={26},
  number={6},
  pages={1--39},
  year={2021},
  publisher={Springer}
}

@inproceedings{mozgovoy2017unity,
  title={Unity application testing automation with appium and image recognition},
  author={Mozgovoy, Maxim and Pyshkin, Evgeny},
  booktitle={International Conference on Tools and Methods for Program Analysis},
  pages={139--150},
  year={2017},
  organization={Springer}
}

@article{petrillo2009went,
  title={What went wrong? A survey of problems in game development},
  author={Petrillo, F{\'a}bio and Pimenta, Marcelo and Trindade, Francisco and Dietrich, Carlos},
  journal={Computers in Entertainment (CIE)},
  volume={7},
  number={1},
  pages={1--22},
  year={2009},
  publisher={ACM New York, NY, USA}
}

@inproceedings{lewis2010went,
  title={What went wrong: a taxonomy of video game bugs},
  author={Lewis, Chris and Whitehead, Jim and Wardrip-Fruin, Noah},
  booktitle={Proceedings of the fifth international conference on the foundations of digital games},
  pages={108--115},
  year={2010}
}

@article{stacey2009temporal,
  title={A temporal perspective of the computer game development process},
  author={Stacey, Patrick and Nandhakumar, Joe},
  journal={Information Systems Journal},
  volume={19},
  number={5},
  pages={479--497},
  year={2009},
  publisher={Wiley Online Library}
}

@inproceedings{murphy2014cowboys,
  title={Cowboys, ankle sprains, and keepers of quality: How is video game development different from software development?},
  author={Murphy-Hill, Emerson and Zimmermann, Thomas and Nagappan, Nachiappan},
  booktitle={Proceedings of the 36th International Conference on Software Engineering},
  pages={1--11},
  year={2014}
}

@inproceedings{politowski2020dataset,
  title={Dataset of video game development problems},
  author={Politowski, Cristiano and Petrillo, Fabio and Ullmann, Gabriel Cavalheiro and de Andrade Werly, Josias and Gu{\'e}h{\'e}neuc, Yann-Ga{\"e}l},
  booktitle={Proceedings of the 17th International Conference on Mining Software Repositories},
  pages={553--557},
  year={2020}
}

@inproceedings{chen2021glib,
  title={{GLIB}: towards automated test oracle for graphically-rich applications},
  author={Chen, Ke and Li, Yufei and Chen, Yingfeng and Fan, Changjie and Hu, Zhipeng and Yang, Wei},
  booktitle={Proceedings of the 29th ACM Joint Meeting on European Software Engineering Conference and Symposium on the Foundations of Software Engineering},
  pages={1093--1104},
  year={2021}
}

@article{davarmanesh2020automating,
  title={Automating Artifact Detection in Video Games},
  author={Davarmanesh, Parmida and Jiang, Kuanhao and Ou, Tingting and Vysogorets, Artem and Ivashkevich, Stanislav and Kiehn, Max and Joshi, Shantanu H and Malaya, Nicholas},
  journal={arXiv preprint arXiv:2011.15103},
  year={2020}
}

@inproceedings{ling2020using,
  title={Using deep convolutional neural networks to detect rendered glitches in video games},
  author={Ling, Carlos and Tollmar, Konrad and Gissl{\'e}n, Linus},
  booktitle={Proceedings of the AAAI Conference on Artificial Intelligence and Interactive Digital Entertainment},
  volume={16},
  number={1},
  pages={66--73},
  year={2020}
}

@inproceedings{nantes2008framework,
  title={A Framework for the Semi-Automatic Testing of Video Games.},
  author={Nantes, Alfredo and Brown, Ross and Maire, Frederic},
  booktitle={AIIDE},
  year={2008}
}

@inproceedings{taesiri2022clip,
        title={{CLIP} meets {G}ame{P}hysics: Towards bug identification in gameplay videos using zero-shot transfer learning}, 
        author={Mohammad Reza Taesiri and Finlay Macklon and Cor-Paul Bezemer},
        booktitle={2022 IEEE/ACM 19th International Conference on Mining Software Repositories (MSR)},
        year={2022},
        organization={IEEE}
  }

@inproceedings{kim2020synthesizing,
  title={Synthesizing Retro Game Screenshot Datasets for Sprite Detection.},
  author={Kim, Chanha and Kim, Jaden and Osborn, Joseph C},
  booktitle={AIIDE Workshops},
  year={2020}
}

@article{smirnov2021marionette,
  title={Marionette: Self-supervised sprite learning},
  author={Smirnov, Dmitriy and Gharbi, Michael and Fisher, Matthew and Guizilini, Vitor and Efros, Alexei and Solomon, Justin M},
  journal={Advances in Neural Information Processing Systems},
  volume={34},
  year={2021}
}

@article{he2015deep,
  title={Deep residual learning for image recognition. arXiv 2015},
  author={He, Kaiming and Zhang, Xiangyu and Ren, Shaoqing and Sun, Jian},
  journal={arXiv preprint arXiv:1512.03385},
  year={2015}
}

@inproceedings{rezende2017malicious,
  title={Malicious software classification using transfer learning of resnet-50 deep neural network},
  author={Rezende, Edmar and Ruppert, Guilherme and Carvalho, Tiago and Ramos, Fabio and De Geus, Paulo},
  booktitle={2017 16th IEEE International Conference on Machine Learning and Applications (ICMLA)},
  pages={1011--1014},
  year={2017},
  organization={IEEE}
}

@inproceedings{chen2020unblind,
  title={Unblind your apps: Predicting natural-language labels for mobile gui components by deep learning},
  author={Chen, Jieshan and Chen, Chunyang and Xing, Zhenchang and Xu, Xiwei and Zhut, Liming and Li, Guoqiang and Wang, Jinshui},
  booktitle={2020 IEEE/ACM 42nd International Conference on Software Engineering (ICSE)},
  pages={322--334},
  year={2020},
  organization={IEEE}
}

@inproceedings{luo2019making,
  title={Making cnns for video parsing accessible: event extraction from dota2 gameplay video using transfer, zero-shot, and network pruning},
  author={Luo, Zijin and Guzdial, Matthew and Riedl, Mark},
  booktitle={Proceedings of the 14th International Conference on the Foundations of Digital Games},
  pages={1--10},
  year={2019}
}

@inproceedings{luo2018player,
  title={Player experience extraction from gameplay video},
  author={Luo, Zijin and Guzdial, Matthew and Liao, Nicholas and Riedl, Mark},
  booktitle={Fourteenth Artificial Intelligence and Interactive Digital Entertainment Conference},
  year={2018}
}

@inproceedings{ye2021empirical,
  title={An empirical study of {GUI} widget detection for industrial mobile games},
  author={Ye, Jiaming and Chen, Ke and Xie, Xiaofei and Ma, Lei and Huang, Ruochen and Chen, Yingfeng and Xue, Yinxing and Zhao, Jianjun},
  booktitle={Proceedings of the 29th ACM Joint Meeting on European Software Engineering Conference and Symposium on the Foundations of Software Engineering},
  pages={1427--1437},
  year={2021}
}

@inproceedings{viggiatousing,
title = {Using Natural Language Processing Techniques to Improve Manual Test Case Descriptions},
author = {Markos Viggiato and Dale Paas and Chris Buzon and Cor-Paul Bezemer},
year = {2022},
date = {2022-05-08},
booktitle = {International Conference on Software Engineering - Software Engineering in Practice (ICSE - SEIP) Track}
}

@article{viggiato2021identifying,
  title={Identifying Similar Test Cases That Are Specified in Natural Language},
  author={Viggiato, Markos and Paas, Dale and Buzon, Chris and Bezemer, Cor-Paul},
  journal={IEEE Transactions on Software Engineering},
  year={2022},
  publisher={IEEE}
}

@inproceedings{zhang2018unreasonable,
  title={The unreasonable effectiveness of deep features as a perceptual metric},
  author={Zhang, Richard and Isola, Phillip and Efros, Alexei A and Shechtman, Eli and Wang, Oliver},
  booktitle={Proceedings of the IEEE conference on computer vision and pattern recognition},
  pages={586--595},
  year={2018}
}

@article{taesiri2020video,
  title={A video game testing method utilizing deep learning},
  author={Taesiri, Mohammad Reza and Habibi, Moslem and Fazli, Mohammad Amin},
  journal={Iran Journal of Computer Science},
  volume={17},
  number={2},
  year={2020}
}

@article{fazli2021under,
  title={Under the Skin of Foundation {NFT} Auctions},
  author={Fazli, MohammadAmin and Owfi, Ali and Taesiri, Mohammad Reza},
  journal={arXiv preprint arXiv:2109.12321},
  year={2021}
}

@article{wang2002universal,
  title={A universal image quality index},
  author={Wang, Zhou and Bovik, Alan C},
  journal={IEEE signal processing letters},
  volume={9},
  number={3},
  pages={81--84},
  year={2002},
  publisher={IEEE}
}

@article{eskicioglu1995image,
  title={Image quality measures and their performance},
  author={Eskicioglu, Ahmet M and Fisher, Paul S},
  journal={IEEE Transactions on communications},
  volume={43},
  number={12},
  pages={2959--2965},
  year={1995},
  publisher={IEEE}
}

@article{wang2004image,
  title={Image quality assessment: from error visibility to structural similarity},
  author={Wang, Zhou and Bovik, Alan C and Sheikh, Hamid R and Simoncelli, Eero P},
  journal={IEEE transactions on image processing},
  volume={13},
  number={4},
  pages={600--612},
  year={2004},
  publisher={IEEE}
}

@article{mann1947test,
  title={On a test of whether one of two random variables is stochastically larger than the other},
  author={Mann, Henry B and Whitney, Donald R},
  journal={The annals of mathematical statistics},
  pages={50--60},
  year={1947},
  publisher={JSTOR}
}

@article{cliff1993dominance,
  title={Dominance statistics: Ordinal analyses to answer ordinal questions.},
  author={Cliff, Norman},
  journal={Psychological bulletin},
  volume={114},
  number={3},
  pages={494},
  year={1993},
  publisher={American Psychological Association}
}

@inproceedings{romano2006exploring,
  title={Exploring methods for evaluating group differences on the NSSE and other surveys: Are the t-test and Cohen’sd indices the most appropriate choices},
  author={Romano, Jeanine and Kromrey, Jeffrey D and Coraggio, Jesse and Skowronek, Jeff and Devine, Linda},
  booktitle={annual meeting of the Southern Association for Institutional Research},
  pages={1--51},
  year={2006},
  organization={Citeseer}
}

\end{document}